\documentclass[twocolumn,prb,reprint,showpacs,preprintnumbers,amsmath,amssymb,superscriptaddress]{revtex4-1}
\usepackage{graphicx}% Include figure files
\usepackage{dcolumn}% Align table columns on decimal point
\usepackage{bm}% bold math
\usepackage{graphicx}
\usepackage{subfigure}
\usepackage{physics}
\usepackage{epstopdf}
\usepackage{braket}
\usepackage{enumerate}
\usepackage{notes2bib}

\newcommand*{\citen}[1]{%
  \begingroup
    \romannumeral-`\x % remove space at the beginning of \setcitestyle
    \setcitestyle{numbers}%
    \cite{#1}%
  \endgroup   
}
\begin{document}

\title{Non-equilibrium Green's function study of magneto-conductance features and oscillations in clean and disordered nanowires}
\author{Aritra Lahiri}
\affiliation{School of Physics and Astronomy, University of Minnesota, Minneapolis, MN 55455, USA\\}
\affiliation{Department of Electrical Engineering, Indian Institute of Technology Bombay, Powai, Mumbai-400076, India\\}
\author{Kaveh Gharavi}
\affiliation{Institute for Quantum Computing, University of Waterloo, Waterloo, Ontario N2L 3G1, Canada}
\affiliation{Department of Physics and Astronomy, University of Waterloo, Waterloo, Ontario N2L 3G1, Canada}
\author{Jonathan Baugh}
\email{baugh@uwaterloo.ca}
\affiliation{Institute for Quantum Computing, University of Waterloo, Waterloo, Ontario N2L 3G1, Canada}
\affiliation{Department of Physics and Astronomy, University of Waterloo, Waterloo, Ontario N2L 3G1, Canada}
\affiliation{Department of Chemistry, University of Waterloo, Waterloo, Ontario N2L 3G1, Canada}
\author{Bhaskaran Muralidharan}
\email{bm@ee.iitb.ac.in}
\affiliation{Department of Electrical Engineering, Indian Institute of Technology Bombay, Powai, Mumbai-400076, India\\}
\date{\today}

\medskip
\widetext
\begin{abstract}
We explore various aspects of magneto-conductance oscillations in semiconductor nanowires, developing quantum transport models based on the non-equilibrium Green's function formalism. In the clean case, Aharonov-Bohm (AB - h/e) oscillations are found to be dominant, contingent upon the surface confinement of electrons in the nanowire. We also numerically study disordered nanowires of finite length, bridging a gap in the existing literature. By varying the nanowire length and disorder strength, we identify the transition where Al'tshuler-Aronov-Spivak (AAS - h/2e) oscillations start dominating, noting the effects of considering an open system. Moreover, we demonstrate how the relative magnitudes of the scattering length and the device dimensions govern the relative dominance of these harmonics with energy, revealing that the AAS oscillations emerge and start dominating from the center of the band, much higher in energy than the conduction band-edge. We also show the ways of suppressing the oscillatory components (AB and AAS) to observe the non-oscillatory weak localization corrections, noting the interplay of scattering, incoherence/dephasing, the geometry of electronic distribution, and orientation of magnetic field. This is followed by a study of surface roughness which shows contrasting effects depending on its strength and type, ranging from magnetic depopulation to strong AAS oscillations. Subsequently, we show that dephasing causes a progressive degradation of the higher harmonics, explaining the re-emergence of the AB component even in long and disordered nanowires. Lastly, we show that our model qualitatively reproduces the experimental magneto-conductance spectrum in [Holloway et al, PRB 91, 045422 (2015)] reasonably well while demonstrating the necessity of spatial-correlations in the disorder potential, and dephasing. 
\end{abstract}
\maketitle
\section{\label{intro}Introduction}
Quantum transport in nanowires is a widely studied topic with numerous applications, many of which are yet to be explored. Nanowires fabricated from narrow band-gap semiconductors harbor surface confined states due to the pinning of Fermi-level over the conduction band edge \cite{PhysRevLett.101.106803,doi:10.1021/nl201102a,PSSC:PSSC200982506}. This results in the formation of a cylindrical two dimensional electron gas (2DEG) on the surface\cite{PhysRevLett.24.303,PhysRevLett.76.3626,PSSC:PSSC200982506}. Cylindrical surface confinement may also be brought about by core-shell heterostructured nanowires \cite{Lauhon2002}. Such a surface distribution over the nanowire forms a multiply connected domain, topologically equivalent to a ring. Transport in closed multiply connected structures, such as rings or cylinders can result in interference between paths having completed different number of loops. Such paths pick up different phase factors in a magnetic field, which leads to oscillations in the conductance with periods in multiples of $h/e$, the Aharonov Bohm (AB) period, depending on the paths involved \cite{PhysRev.115.485,PhysRevLett.114.076802}.

Experiments performed on core shell nanowires \cite{RichterNanoLett2008} and nanowires with surface confined electronic distributions \cite{PhysRevB.91.045422,PhysRevB.77.075332,PSSC:PSSC200982506} have revealed AB oscillations. Appreciable and sustained oscillations over a large range of applied fields are however contingent upon the presence of a superficial conduction layer. Hence the magneto-conductance traces can be used to probe the presence of such surface confined states. Such experiments and theoretical studies have also been performed on topological insulator nanowires \cite{Peng2010,PhysRevLett.110.186806,arxivtiripple,PhysRevB.97.035157}. Moreover, superconductor-semiconductor hybrid nanowire devices are used to study Majorana fermions where a study of magneto-conductance and disorder is critical~\cite{PhysRevB.85.064512}. Further, the conductance of disordered nanowires have revealed the Al'tshuler-Aronov-Spivak (AAS) oscillations\cite{Sharvinsharvin} with period $h/2e$ . Given various theoretical and experimental endeavors on magneto-conductance of nanowires \cite{AIPwashburn,RevModPhys.59.755,EPLexp,PhysRevB.89.045417}, there is still a lack of a comprehensive theoretical study of various aspects of magneto-transport in both clean and disordered nanowires. In this paper, we try to bridge this gap and extend previous studies by employing non-equilibrium Green's function (NEGF) simulations in clean and disordered nanowires.

This paper is organized as follows. We begin by providing a brief review of the origin of oscillations in clean and disordered individual nanowires with axial magnetic field. This is followed by a study of clean nanowires in the clean and ballistic limit, where we highlight the role of surface confinement in producing the AB oscillations. However, in reality, semiconductor nanowires are plagued by various sources of disorder such as short-range unscreened potential impurities and surface roughness, as well as dephasing/decoherence. Accordingly, by considering a range of nanowire lengths and disorder strengths, we show a transition point where the AAS oscillations start dominating over the AB oscillations and thus bridge the gap between the clean and the disordered limits. Further, it is found that the oscillatory behavior depends on the position of the Fermi level, dividing the band into distinct regions each with a dominant type of oscillation, depending on the dispersion of the surface states and the constituent material. This analysis is performed by comparing the scattering length with the device dimension, which provides a guideline to predict the oscillatory behavior. Next, we highlight the conditions for observing the non-oscillatory weak localization corrections. We then study the effects of surface roughness and dephasing on the transmission spectrum. We find that while surface roughness may again lead to the dominance of AAS oscillations, dephasing systematically degrades the oscillations, starting with the higher harmonics which can explain the experimentally observed AB oscillations in disordered nanowires. Lastly, we investigate the previously studied effects in nanowires with a parabolic transverse potential leading to a weaker surface confinement of electrons. Our model produces a reasonably well qualitative reproduction of the experimental magneto-conductance spectrum in Ref.~\citen{PhysRevB.91.045422},  highlighting the interplay of spatially-correlated disorder scattering and dephasing  to reproduce the experimentally observed features.

\section{\label{surfc}Surface conduction in a nanowire}
In this section we give a brief theoretical overview of the physics behind AB and AAS oscillations and their effects on the conductance of a nanowire, using simple steady state equations. We consider a nanowire subjected to an axial magnetic field, in which the electronic motion may be broken down into two parts: the axial motion along the length of the nanowire, and the motion along the circumferential direction. While traversing the nanowire axially, electrons moving along the circumferential direction interfere, resulting in a number of effects which shall be explored in this work. For the simple case of electrons confined only on the surface of the nanowire, thereby behaving as an infinitely long cylindrical system of radius $R$, the Hamiltonian is given by $H = \frac{\hbar^2}{2m}\frac{d^2}{dz^2}+\frac{1}{2m}\left(\frac{-i\hbar}{R}\frac{\partial}{\partial \phi}-\frac{e\Phi}{2\pi R}\right)^2$. The eigen-energies $E(l)$ satisfy
\begin{align}
E(l) = \frac{\hbar^2k_z^2}{2m^*}+\frac{\hbar^2}{2m^*R^2}\left( l-\frac{\Phi}{\Phi_0}\right)^2,
\label{leval}
\end{align}
where $k_z$ represents the wave number along the $\hat{z}-$direction, $m^{*}$ is the effective mass, $\Phi=\pi R^2B$ is the applied axial magnetic flux , with an axial magnetic field ($B$) along z-axis, $\Phi_0 = \frac{h}{e}$-the magnetic flux quantum, $e$ is the electronic charge and $h$ is the Planck's constant. While traversing the nanowire axially, electrons move on the surface along series of rings, and interfere. Quantum corrections to conductivity due to the interference between two time-reversed trajectories have been explored previously \cite{JETP.35.11.588,PhysRevB.89.045417}.
The conductance can be evaluated from the phase factors arising from the interference terms, which may be calculated using path integrals by summing over the factors corresponding to paths having given winding numbers
\begin{align}
&\mathcal{K}(\theta',t',\theta,t)=\braket{\Theta(t')|\Theta(t)}=\sum_n\int_{C_n}d\theta e^{S(\theta,\mathbf{A})/\hbar}\nonumber\\
&=\mathrm{exp} \left( i e\Phi(\theta'-\theta)/h \right)\left(\sum_n e^{i2\pi ne\Phi/h}\mathcal{K}_n \left( \theta',t',\theta, t \right)\right), \label{propagator}
\end{align}
where $\mathcal{K}$ and $\mathcal{K}_n$ are the full propagator, and propagators (without magnetic field, as the magnetic part has already been extracted out) restricted to paths having a winding number $n$($C_n$) respectively, and $S(\theta,\mathbf{A})$ is the action.
From the Landauer formalism, the conductance may be written as $\mathcal{G}(E)=\frac{2e^2}{h}\lvert\mathcal{K}(E)\rvert^2$. Now all such paths given by the terms inside the bracket in~\eqref{propagator} interfere. For example, paths with winding numbers $n$ and $-n$ combine to result in an oscillation of amplitude $ \alpha_n \mathrm{cos}(4\pi n e/h)$ having a period $h/2ne$. The dynamics of such a situation can be captured in the expression for the propagator. The conductivity correction can be given by, $\Delta \sigma (H)/\sigma_0 = \beta \left(\sum_n \alpha_n \mathrm{cos}(2\pi n e/h) \right) $, where $\alpha_n$ and $\beta$ are constants~\cite{PhysRevB.74.245327}.

This can also be seen in a different way in steady state. In the absence of a flux through the nanowire, the angular part of the solutions $(\Psi(r_0,\theta))$ form sinusoids on the surface for a given angular momentum quantum number $l$, along with an axially propagating component. Assuming in general that a state is a linear combination of mutually orthonormal angular momentum eigenstates, the norm of the angular part is given by
\begin{align}
\Psi(\phi) &= \sum_{l=-\infty}^{l=\infty}r_l\mathrm{exp} (il\phi).
\end{align}
In the presence of a magnetic field, the rotating states pick up a phase, dependent on the angular momentum eigenstate considered. Assuming that a wave-packet incident from one of the leads has had sufficient time to acquire steady state in its angular part, at any angle $\phi$ along the circumference, the wave-function is written as a sum of components having different winding numbers(n). However such a description can lead to multi-valued wave functions on the application of magnetic field (including the Peierl's phase factor) for non-integral values of $\Phi/\Phi_0$, as the phase acquired over a loop is not necessarily $2\pi$.

This technicality can however be avoided by switching into a description of the steady state solution in terms of waves with given winding numbers using the Poisson summation formula, drawing from the discussion by Berry~\cite{0143-0807-1-4-011}. This describes the wave at a point as a sum of waves arriving at that point after traversing different loop numbers:
\begin{align}
\Psi(\phi) &= \sum_{n=-\infty}^{\infty}W_n(\phi), \\
W_n(\phi) &= \int_{-\infty}^{\infty}  r(\gamma)\mathrm{exp} (i\gamma\phi)\mathrm{exp} (i2\pi n\gamma) \mathop{ \mathrm {d} \gamma},
\end{align}
where $W_n$ is a wave function with winding number n, and $\gamma$ interpolates $l$ to non-integral values. On assigning the Peierls' phase factor accordingly,
\begin{align}
\Psi(\phi) &= \sum_{n=-\infty}^{\infty}W_n(\phi)\mathrm{exp}\left( i(2\pi n+\phi)\frac{\Phi}{\Phi_0}\right), \label{windingdecomp}
\end{align}
which avoids the problem above since
\begin{multline}
\Psi(\phi+2\pi) = \sum_{n=-\infty}^{\infty}W_n(\phi+2\pi)\mathrm{exp}\left( i2\pi (n+1 + \phi)\frac{\Phi}{\Phi_0}\right)  \\
 = \sum_{n=-\infty}^{\infty}W_{n+1}(\phi)\mathrm{exp}\left( i2\pi (n+1 + \phi)\frac{\Phi}{\Phi_0}\right) =\Psi(\phi). 
\end{multline}
Note that after a revolution, the weight of $W_n$ is transferred to $W_{n+1}$, as $W_n(\phi+2\pi)=W_{n+1}(\phi)$.

Now, in \eqref{windingdecomp}, the terms in the summation interfere due to the presence of the phase factors, giving rise to the oscillatory behavior with respect to the magnetic flux $\Phi$. Note that, for the disordered case, the phase picked by by $W_n(\phi)$ is now $\mathrm{exp}\left( i2\pi (n+1 + \phi)\frac{\Phi}{\Phi_0}\right)\mathrm{exp}(ig(n,\phi))$. Here the first exponential term is the magnetic phase factor, and the second exponential term is the disorder potential induced phase factor.  Here $g(n,\phi)$ is the classical action for the corresponding path without the magnetic field. Also, $g(n,\phi)$ is real for our case of real disorder potential. Now $\lvert \Psi(\phi) \rvert ^2$ has oscillatory components corresponding to the terms (for the general disordered case),
\begin{align}
\lvert W_x(\phi) \rvert \lvert W_y(\phi) \rvert \mathrm{cos}( f(x+y)) \mathrm{cos}\left( 2\pi (\lvert x-y\rvert)\frac{\Phi}{\Phi_0}\right), \label{osc_har}
\end{align}
where $f(x+y)=g^*(x,\phi)-g(y,\phi)$ is the path dependent random phase, which evaluates to zero when $y=-x$, and is dependent on the disorder configuration along the path traversed by the states labeled by $x,y$. The path dependent phases are phase factors picked up by rotating waves in the absence of a magnetic field, and in general depend on energy and disorder. However for paths $x$ and $y$, satisfying $y=-x$, the phases cancel, as mentioned earlier. Such paths contribute to oscillations with period $h/me$ in flux with even $m$. Higher harmonics are present too, but with much smaller magnitude corresponding to higher winding numbers or longer paths traversed.

The AB oscillations can also be connected to the eigenvalue spectrum of a ring, which is a series of parabolas as noted in \eqref{leval}, with their minima on integral values of $\Phi/\Phi_0$. The conductance changes each time a parabola crosses the Fermi level, while the flux is increased. The origin of higher harmonics however cannot be explained by this simple picture.

From \eqref{osc_har}, it is seen that the first oscillatory component has a period $\frac{h}{e}$. Note that this component arises from paths with path difference equal to the circumference of the nanowire. This can happen when $(x,y)=\{(1,0)$ $(2,1)\ldots\}$ in ~\eqref{osc_har}. Further, one may also introduce a shift of $\pi$ to the final (=$\theta'$ in~\eqref{propagator}) angular position of both the interfering paths relative to the initial (=$\theta$ in~\eqref{propagator}) angular position, such that one may effectively have a winding number $(x,y)=\{(1/2,-1/2)$. Thermal averaging in a suitable energy interval at non zero temperatures and/or ensemble averaging in the presence of multiple parallely connected rings in a cylinder can dampen the $\frac{h}{e}$ oscillations. The next significant mode observed is the $\frac{h}{2e}$ periodic oscillation. The primary contributor to this harmonic corresponds to $(x,y)=(1,-1)/(-1,1)$, which are independent of any random phase, and are consequently robust against disorder (elastic scattering) and other fluctuations. Clearly, the paths involved in this harmonic enclose twice the amount of flux as the paths involved in the $h/e$ harmonic.

As seen from~\eqref{osc_har}, all oscillations with period $h/ne$, where $n$ is even (formed by $x=n$ and $y=-n$) are independent of random phases, making them theoretically resistant to disorder and thermal averaging, within the scope of this analysis. Ideally, in thin rings, oscillations with period $h/e$ dominate, with higher harmonics appearing with smaller amplitudes. However in disordered rings with finite widths, $h/2e$ periodic oscillations become survive and a crossover occurs where $h/2e$ becomes the dominant component \cite{RevModPhys.59.755}. Such behaviour is also expected in cylindrical conductors, either due to impurity based disorder or surface roughness and radial randomness along the cylinder axis. We now proceed to a numerical analysis of the aforesaid features.
\section{RESULTS}
\subsection{\label{model}TIGHT BINDING NEGF MODEL}
\begin{figure}[htb!]
\subfigure[]{\includegraphics[width=1.6in]{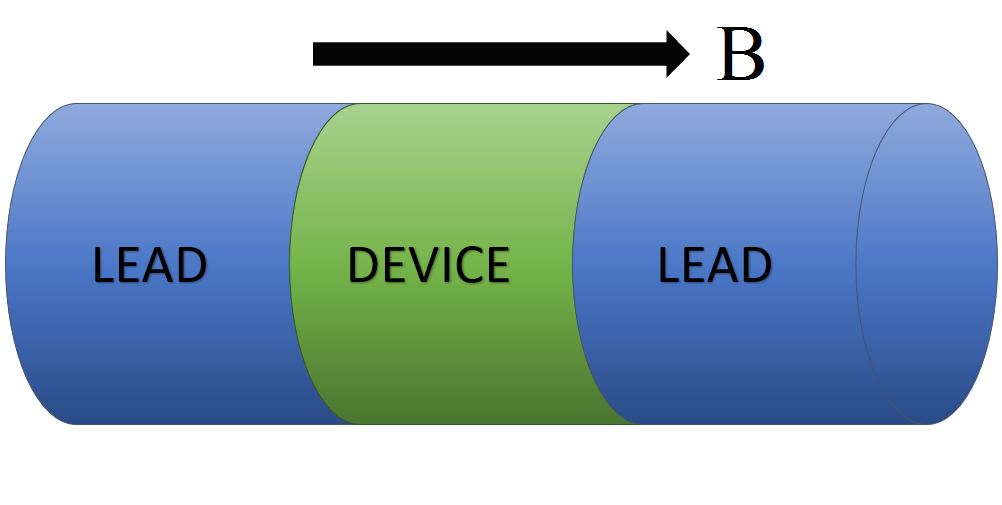}}
\hspace{1cm}	
\subfigure[]{\includegraphics[width=0.95in]{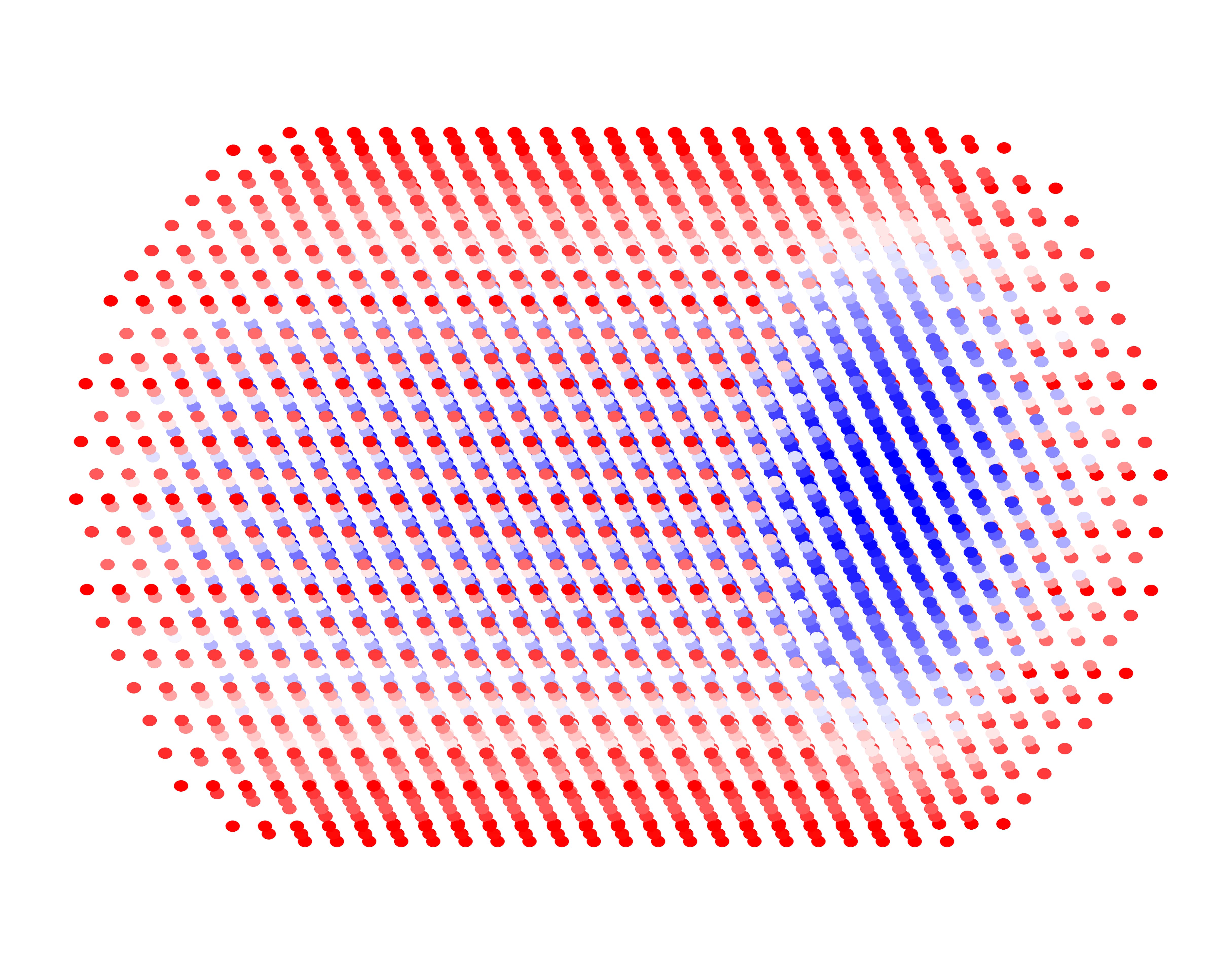}}
\caption{Device schematics: (a) The nanowire device and the leads, along with an axial magnetic field. (b) Cubic lattice structure of the nanowire with a surface confining potential. The region marked in red has lower electronic energy than the region marked in blue. This results in a surface confined electronic distribution (in the red region).}
\label{device_design}
\end{figure}

In this section we introduce the system and the non equilibrium Green's function (NEGF) formalism of quantum transport being used in this work. We consider a nanowire with a cubic lattice as shown in Fig.~\ref{device_design}. To capture the essential physics, we use a tight binding Hamiltonian with a single basis orbital for each lattice-site.
\begin{equation}
\mathcal{H} =  \sum_{\langle i, j \rangle}\left( t\hat{c}^\dagger_{i}\hat{c}^{\phantom{\ast}}_{j} + (\epsilon + U_{i})\hat{c}^\dagger_{i}\hat{c}^{\phantom{\ast}}_{i} \right),
\label{TB H}
\end{equation}
where $\epsilon$ is the on-site energy, $t$ is the hopping parameter, $U_i$ is the random potential at each site, and the sum is over nearest neighbours $i,j$ (good approximation in the maximally localized Wannier basis). The operators $\hat{c}^{\dagger}_{i}\hspace{0.1cm}(\hat{c}^{\phantom{\ast}}_{i})$ represent the creation (annihilation) operators for electrons at site $i$. 
For obtaining a cross-section of our choice, in our case a disc, suitable potential has been added to simulate the band offset of the required geometry. A very high band offset works as well as hard wall boundary conditions implemented by brute force shaping of the Hamiltonian. Note that cylindrical symmetry has been assumed throughout the study. In the presence of an external vector potential \textbf{A}, the hopping parameters between sites $m$ and $n$ acquire a Peierl's phase\cite{Peierls1933}, 
\begin{equation}
t_{mn} \rightarrow t_{mn}e^{i2\pi\frac{e}{\hbar}\int_{\mathbf{r_n}}^{{\mathbf{r_m}}}d\mathbf{r}\cdot\mathbf{A(r,}t)}=t_{mn}e^{2\pi\frac{\Phi}{\Phi_0}}.
\end{equation}
Transport calculations are performed using the NEGF/Keldysh formalism\cite{keldyshjetp} as discussed in ref. \cite{dattablack}. We begin with the one particle retarded and lesser Green's function
\begin{align}
G^R(\mathbf{x},t;\mathbf{x},t')&=-i\Theta(t-t')\langle \{\hat{c}(\mathbf{x},t),\hat{c}^\dagger(\mathbf{x}',t')\}\rangle,\\ 
G^<(\mathbf{x},t;\mathbf{x},t')&=i\langle\hat{c}^\dagger(\mathbf{x},t),\hat{c}(\mathbf{x}',t')\rangle,
\end{align}
where $\mathbf{x},\mathbf{x}'$ represent the initial and the final states respectively, on the standard time contour. The advanced Green's function is given by $G^A=(G^R)^\dagger$.

In the non-equilibrium steady state, it admits an energy domain representation after Fourier transforming. In the energy domain real space matrix representation, we have,
\begin{align}
G^R(E) &= \left[(E+i\eta)I-H_0-U-\Sigma(E)\right]^{-1},
\end{align}
where $\eta$ is an infinitesimal quantity, $E$ is the Fermi energy (controlled by gate voltage), $I$ is an identity matrix of the same size as the system Hamiltonian $H_0$, $U$ is the on-site Anderson disorder potential representing unscreened short-range impurity potentials. In this work $U$ serves the following purposes: a) It adds a band offset to form a cylinder/disc for the nanowire geometry, b) it adds a confining potential to confine electrons on the surface, and c) in disordered nanowires, it adds an on-site scattering potential. The Keldysh formalism is capable of including various scattering mechanisms such as electron-electron, electron-phonon (which can be used for dephasing), disorder averaged quantities etc, through suitable self-energy operators. The net self-energy $\Sigma(E)$ accounts for the leads, and may additionally dress the electrons with appropriate interactions. The implementation of the lead self-energy is detailed in Appendix~\ref{selfE}.

Since we are concerned only with the physics of the system in the limit of very small applied bias and not directly on the device performance, for transmission calculations we assume that the bands are linearly dropping in the axial direction with vanishing slope. Transverse potential is introduced without any self-consistent calculations.

The current for the non-interacting case is given by~\cite{PhysRevLett.68.2512}, 
\begin{align}
J = \frac{e}{h}\int dE(f_L(E)-f_R(E)) \mathbf{Tr}[\Gamma_L(E)G(E)\Gamma_RG^{\dagger}(E)],
\end{align}
where $\Gamma_{L/R}$ represent the tunnel coupling to the leads, which are related to the broadening induced by the contact self-energies ($\Gamma_{L/R} = i(\Sigma_{L/R}-\Sigma_{L/R}^{\dagger})$). The quantity $\mathbf{Tr} [\Gamma_L(E)G(E)\Gamma_RG^{\dagger}(E)]$ is the transmission at the energy $E$, denoted by $T(E)$. Now, the conductance through a level of degeneracy $g$ ($g=2$ for a spin-degenerate level) in the limiting case of vanishing applied voltage and temperature is given by $\mathcal{G} = g\mathcal{G}_0\int dE T(E)(f_L(E)-f_R(E))\rvert_{\substack{eV,T=0}}=g\mathcal{G}_0T(\mu)$, with $\mu$ being the equilibrium Fermi level and $\mathcal{G}_0=e^2/h$. 

In the subsequent sections, the lattice constant is denoted by $a$. Further, all energies and potentials are specified as a scale-invariant quantity, in terms of the hopping parameter $t$. 
\subsection{\label{clean}NON-DISORDERED NANOWIRES}
The presence of an axial magnetic field couples orbital angular momentum states with the field, which causes levels with adjacent angular momentum quantum numbers to shift by one flux quantum. The resulting spectrum is quasi-parabolic and quasi-periodic in nature for small applied fields and can be used to study the sub-band structure. The transmission spectrum is influenced by the electronic distribution, which in turn depends upon the radial confining potential\cite{PhysRevB.91.045422}. Lower surface confinement is found to impart a lower degree of periodicity to the transmission spectrum. This enables a better identification of the sub-bands. Note that throughout the study, the nanowire is assumed to be in the phase coherent regime $(L\leq l_{\phi})$ unless mentioned otherwise, with $l_{\phi}$ being the phase coherence length. Our clean nanowires have a diameter $11a$. Typical experiments have reported phase coherence lengths of the order of a few hundred nanometers\cite{doi:10.1021/nl201102a,PhysRevB.91.045422}, which is longer than the nanowire dimensions considered here. 
\subsubsection{No surface confinement}
\begin{figure}[htb!]
\includegraphics[width=3.5in]{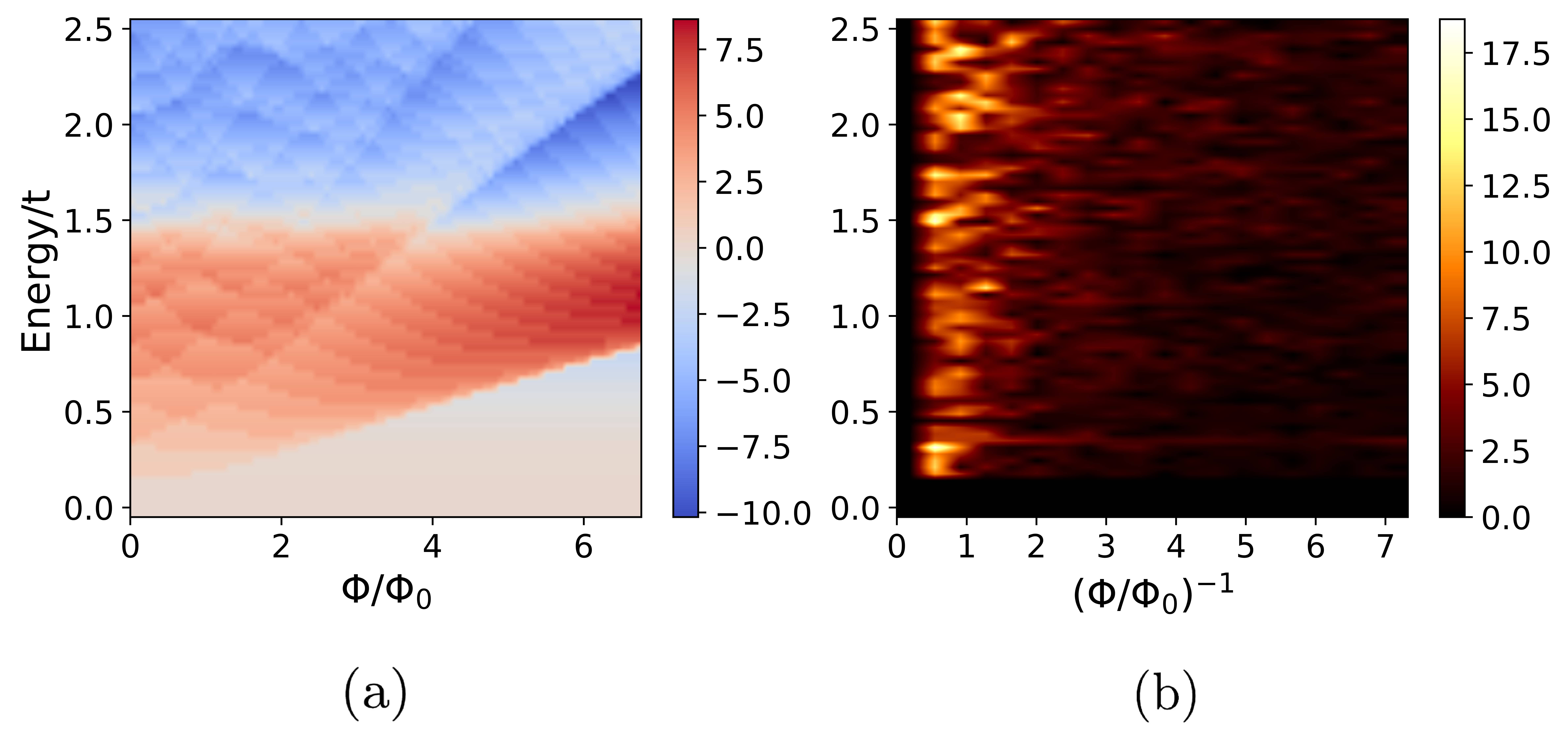}
\caption{The unconfined case: (a) Variation of the transmission, from the mean mean value for each energy, $\delta T(E,\Phi)=T(E,\Phi)-\langle T(E,\Phi) \rangle_\Phi$. No AB oscillations are observed. At small fluxes, it resembles the Fock Darwin spectrum, which quickly aligns with the Landau levels. (b) FFT spectrum of transmission for the same device, which again shows the lack of AB oscillations. Instead we have a broader peak, with frequencies smaller than that of the AB oscillations ($(\Phi/\Phi_0)^{-1}=1$).}
\label{T FFT no pot circstep}
\end{figure}

In this case, the transverse potential is zero within the boundaries of the nanowire, i.e. $V(r)=0$. The plot of the variations of the transmission from its mean (w.r.t. flux) as a function of energy and applied field and its fast fourier transform (FFT) spectrum are depicted in Fig.~\ref{T FFT no pot circstep}(a) and (b). Note that in the quantity $\delta T(E,\Phi)=T(E,\Phi)-\langle T(E,\Phi) \rangle_\Phi$, the average  over flux has been performed over the range shown in panel (a). This holds for all the figures in this work. The transmission resembles the Fock-Darwin spectrum in quantum dots\cite{0953-8984-26-9-095501}, with the transmission increasing in steps at the onset of each level. At higher magnetic fields they however tend to align with the Landau levels\cite{acsnanolett6b00414,0953-8984-26-9-095501}. Subsequent minima move higher in energy as they represent states with a higher orbital angular momentum, which are confined closer to the surface.  The period of the pseudo-oscillations are however greater than one flux quantum ($\Phi_0=h/e$). This happens because the electronic distribution is not confined to the surface. Consequently, electronic paths enclose a smaller flux than the flux through the entire nanowire cross-section. Also, there is no common flux-ratio ($\Phi/\Phi_0$) that may be defined for all states, as different paths can enclose different fluxes. This explains the breadth of the peak. 

\subsubsection{With surface confinement}
\begin{figure}[htb!]	
\includegraphics[width=3.5in]{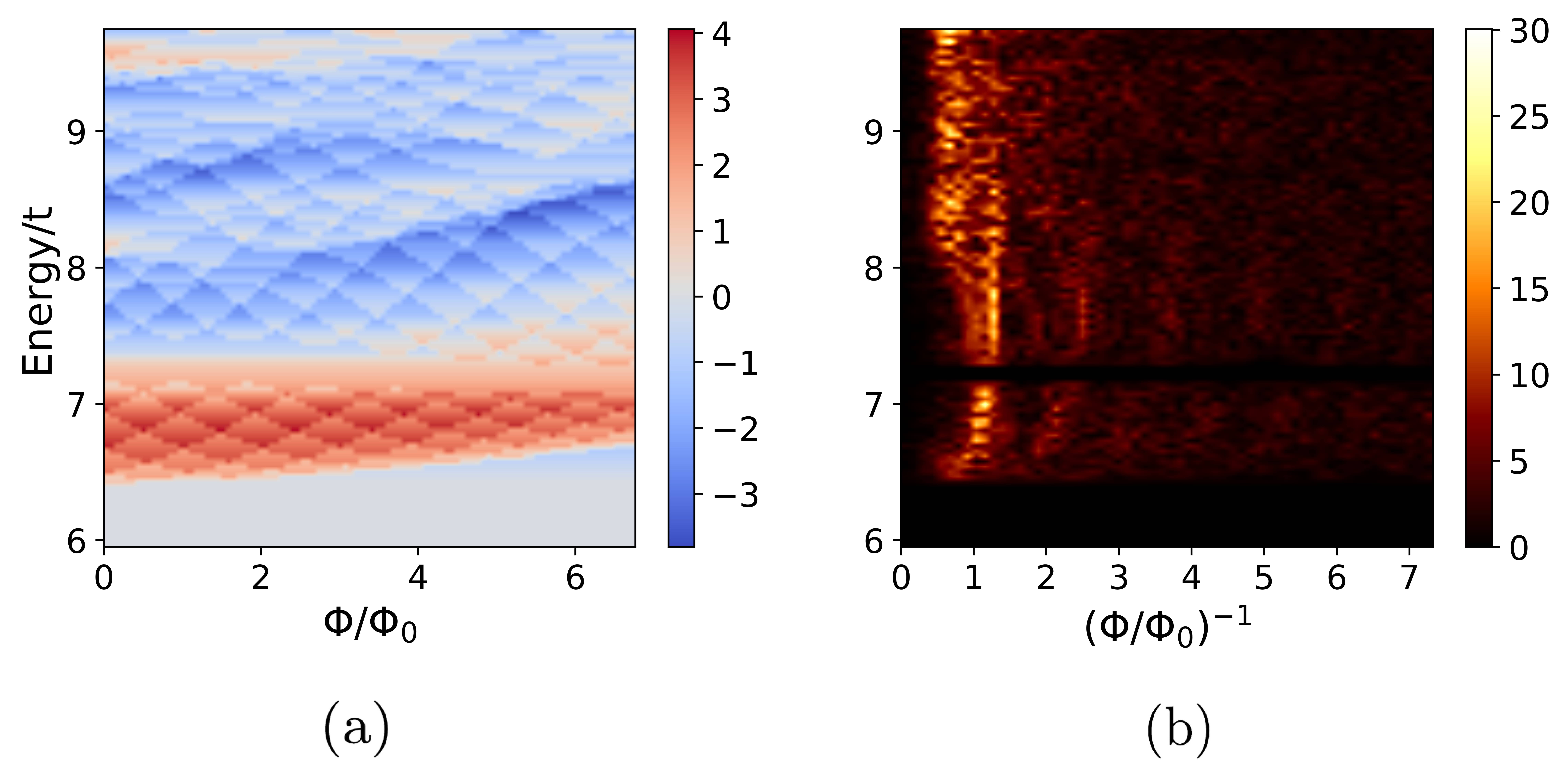}
\caption{The surface confined case with $V_0=0.358t$ and $p=2$: (a) Variation of the transmission, from the mean value (w.r.t flux) for each energy, with a parabolic radial potential. Diamond shaped structures with a period of $h/e$, synonymous with AB oscillations, can be observed at lower energies. At higher values of flux (roughly over $10\Phi_0$), they start moving up, aligning with Landau levels. (b) FFT spectrum of the transmission, showing a peak at $(\Phi/\Phi_0)^{-1}=1$, i.e. the AB peak, at low energies. At higher energies, it gives way to lower frequency components. }
\label{T FFT truepar circstep}
\end{figure}

Now, we add a parabolic transverse potential, to study the effect of surface confinement on the conductance harmonics. The potential, $V(r)$, has the following form,
\begin{equation}
U(r)=-eV(r) = -V_0\left(\frac{r}{R}\right)^p, \label{tVprop}
\end{equation}
where $R$ is the cylinder radius, $U(r)$ is the electronic energy, and the parameters $V_0$ and $p$ are adjusted to ensure strong surface confinement. In Fig.~\ref{T FFT truepar circstep}, we have plotted $\delta T(E,\Phi)=T(E,\Phi)-\langle T(E,\Phi) \rangle_\Phi$ for a nanwire with a transverse potential described by~\eqref{tVprop} with $V_0=0.362t$ and $p=2$, showing the variation of the transmission from its mean value (w.r.t. flux). As seen from Fig.~\ref{T FFT truepar circstep}(a) and (b), as the field varies, more elaborate and sustained diamonds are observed in comparison to the case with no confining potential. Energy levels do not align with the Landau levels as quickly as in the non-confined case, highlighting the effect of surface confinement. The presence of a confining potential invariably confines all states near the surface, which diminishes the difference between the higher and the lower angular momentum states, as the effective radius of the distribution is forced to be the same for all angular momentum states as in Eq.~\ref{leval}. Therefore, the spectrum more or less lies in the same band of energy. Further, proper oscillations are observed, with a strong peak at a frequency corresponding to period $\frac{h}{e}$.

\indent It is observed from Fig.~\ref{T FFT truepar circstep}(b) that at higher energies we get smaller frequency (larger period) oscillations, similar to the unconfined case. This is because at higher energies there exist states which lie away from the surface, i.e., in the bulk. It is a result of using a parabolic confining potential which has a minima at the center of the nanowire, with the electronic energy being highest at the center. These states, with high energy, enclose a smaller flux and hence oscillate with larger periods. 

\begin{figure}[htb!]	
\includegraphics[width=3.3in]{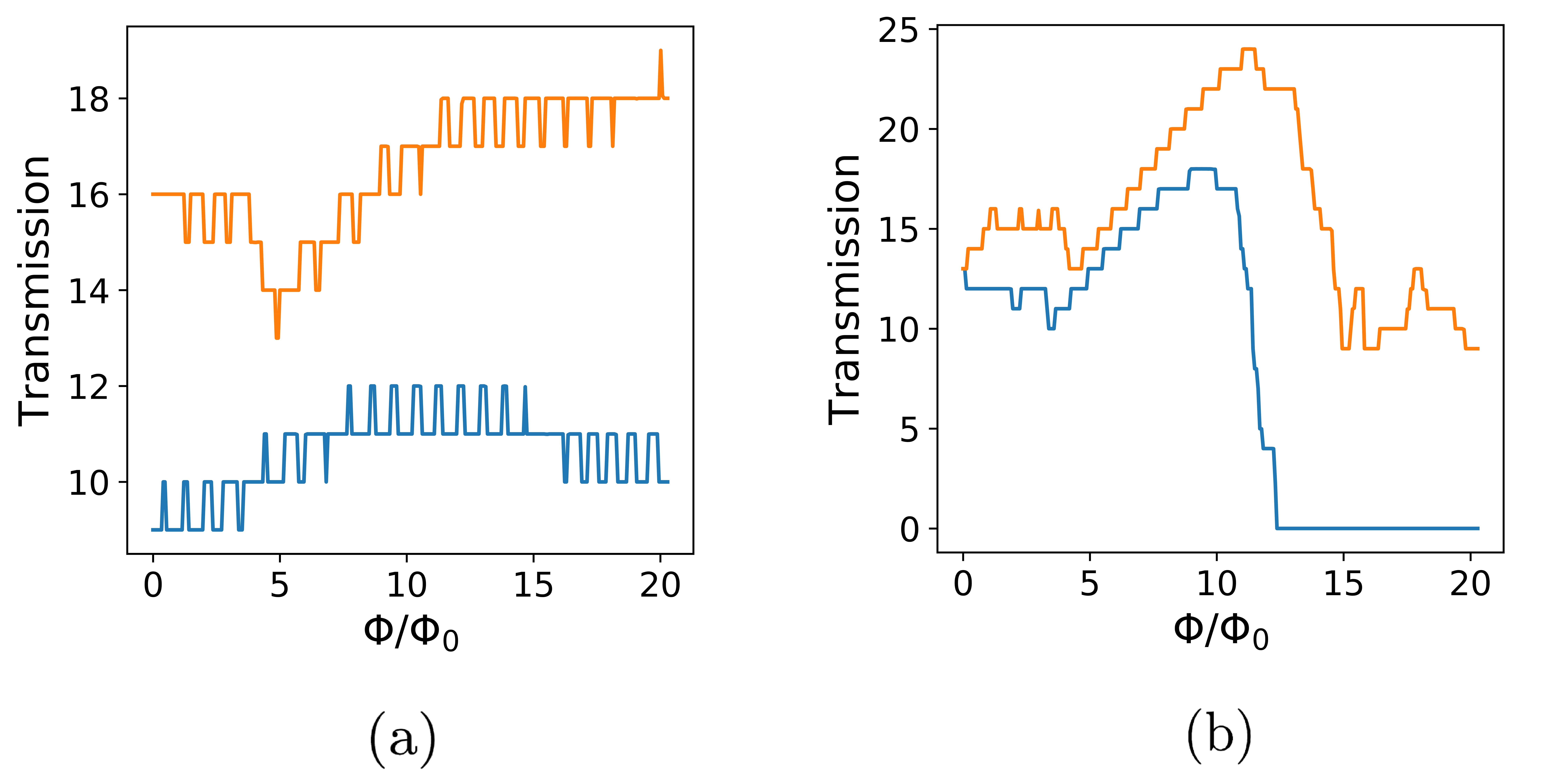}
\caption{Transmission traces/cuts at two different energies(chosen for illustration), with (a) a parabolic surface confining potential, and (b) no surface confining potential, corresponding to Figs.~\ref{T FFT no pot circstep} and~\ref{T FFT truepar circstep}. The orange curves are taken at $E-E_c=1.9t$, and the blue curves are taken at $E-E_c=1.2t$, where $E_c$ is the lower band-edge. Much better and sustained oscillations, with the AB period, are observed in the surface confined case shown in (a). The sharp steps are a consequence of using zero temperature. In the unconfined case shown in (b), sudden drops in conductance in the unconfined case arise due to the passing of energy bands over the Fermi energy, as they align with the Landau levels. Note that the boundary where this depopulation occurs for the unconfined case, follows the same shape as seen in Fig.~\ref{T FFT no pot circstep}, where the transmission drops to zero.}
\label{G traces}
\end{figure}

Note that the cylindrical symmetry of the nanowire under the influence of the gate potential is crucial for observing good AB oscillations. In its absence, a gate potential which depends upon the azimuth angle would suppress paths traversing the section with higher electron energy, and therefore suppress the oscillations. However, for standard oxide thicknesses, such effects are expected to be minor for thin nanowires~\cite{PhysRevB.91.045422}.

In Fig.~\ref{G traces} we show the transmission at two energy values (chosen just for illustration) for both the confined and unconfined case respectively. In the unconfined case, we initially observe irregular variations with a period larger than the AB period. This is followed by sudden drops in the transmission, as all the levels align with the Landau levels at high flux, and move up, over the Fermi energy, depopulating those states. Higher surface confinement pushes the point where alignment begins to higher values of flux. 

\subsection{\label{dis}DISORDERED NANOWIRES}
In the weak disorder regime, low temperature conductivity is largely dominated by elastic scattering via impurity potentials. When the system is of comparable size, or smaller than the phase coherence length, transmission is affected by interference between paths. This results in oscillatory behaviour in the weak-localization corrections in multiply connected systems, in addition to aperiodic, noise-like universal conductance fluctuations (UCF)\cite{PhysRevB.35.1039}. Unlike clean nanowires, in long and weakly disordered nanowires, AB (h/e) component no longer dominates over the AAS (h/2e) component \cite{JETP.35.11.588,PSSC:PSSC200982506}, which has been revealed in experiments too\cite{PSSC:PSSC200982506}\cite{EPLexp}. As mentioned earlier, AB (h/e) oscillations have a random non-magnetic phase, whereas in AAS (h/2e) oscillations, the time-reversed paths show a cancellation of this non-magnetic phase. This occurs due to the paths contributing to AB oscillations facing different environments due to the presence of different disorder configurations. This does not occur in AAS oscillations as both the time reversed paths traverse the same classical path, accumulating the same phase which gets canceled, as seen in~\eqref{windingdecomp} and the discussion following it. The phase change introduced by such rigid elastic scatterers is definite, unlike the phase randomization considered in Sec.~\ref{dephasing}, as might be expected from electron-electron interactions or exchange with a bath as seen in phonon scattering. Simply stated, here we are concerned with the quantum diffusive regime $(l_\phi>l_e)$, whereas in Sec.~\ref{dephasing}, we move towards the classical diffusive regime $(l_\phi<l_e)$. In this study, we have neglected effects of spin orbit interaction (SOI), as for nanowires with small diameters (as has been considered in our study) the spin-relaxation length $l_{\mathrm{SO}}$ has been found to be much larger than the nanowire length/phase coherence length in experiments conducted in InAs and Ga$_\mathrm{x}$In$_{1-\mathrm{x}}$As/InP nanowire\cite{PhysRevB.74.081301,PhysRevB.81.155449}. This is manifested as the appearance of WL corrections instead of weak anti-localization (WAL). Further, Rashba SOI would split the degenerate bands and electrons in each spin-polarized band behave independently of the other band~\cite{weperen}.

The quantum corrections to the conductivity $(\delta\sigma)$ (from the Kubo formula), in diagrammatic terms, are given by the sum of the diffusons (ladder diagrams), and Cooperons (maximally crossed diagrams), which add to the Drude part. For isotropic scatterers in closed systems, the contributions of the ladder diagrams to the conductivity vanish. For a cylindrical electronic distribution penetrated by an axial magnetic field ($\mathbf{B}=\nabla \times \mathbf{A}$), $\delta\sigma$ can be obtained by solving the Cooperon equation\cite{RevModPhys.59.755},
\begin{align}
\left[-i\omega-D\left(\nabla-\frac{2ie}{\hbar}\mathbf{A}\right)^2+\frac{1}{\tau_\phi}\right]C_\omega(\mathbf{r},\mathbf{r}')&=\frac{1}{\tau}\delta(\mathbf{r}-\mathbf{r}'),
\end{align}
which gives us the return probability. For a cylinder of radius $R$, the DC correction is given by\cite{RevModPhys.59.755},
\begin{align}
\delta \sigma = -\frac{e^2}{\pi\hbar} \Bigg[\mathrm{ln}&\left(\frac{l\phi}{l_e}\right)\nonumber \\
&+2\sum_{n=1}^\infty \mathrm{K}_0\left(n\frac{2\pi R}{l_\phi}\right)\mathrm{cos}\left(2\pi n\frac{2\Phi}{\Phi_0}\right)\Bigg], \label{coopcondcorr}
\end{align} 
where $l_\phi=\sqrt{D\tau_\phi}$ is the phase coherence length, $l_e=\sqrt{D\tau}$ is the elastic scattering mean free path, and $\mathrm{K}_0(z=2n\pi R/l_\phi)$ is the Macdonald function, representing the magnitude of the harmonic with period $h/(2ne)$.

While analytical theory exists for closed systems in the clean, and the extreme case of complete disorder averaging, real experiments involve finite nanowires connected to metallic leads. We numerically bridge this gap by studying this transition region with finite disordered nanowires. We look at the quantum diffusion regime, where the device dimensions are larger than the mean free path, but smaller than the phase coherence length. The magnitude of the AAS component in the transmission can be used to probe the degree of disorder. However, the application of disorder of large magnitude puts the nanowire into the Anderson localized regime, where the conductance drops exponentially. In a weakly disordered mesoscopic ring (one specific sample with a specific lead configuration), both AB and AAS oscillations are expected. On taking an ensemble average over multiple rings, this phase causes the AB oscillation to die down. However, a weakly disordered nanowire is in itself an ensemble of multiple weakly disordered rings with longitudinal/axial nanowire slices. This suggests a significant AAS component in sufficiently long disordered nanowires. On the other hand, too short a nanowire would permit electrons to leave before they are able to traverse the required path lengths along the circumference to generate the AAS oscillation. This should allow us to identify a transition point where AAS magnitude becomes larger than AB magnitude. Considering both the effect of disorder strength as well as the role of nanowire length in dictating the magnitudes of the oscillation harmonics, we now present conductance traces and corresponding spectra for both cases.

We consider long disordered nanowires with a strong/sharp surface confining transverse potential which confines the electrons to one layer on the surface. This permits us to explore the essential physics and simulate long nanowires. We consider nanowires with diameter $10a$.
\begin{figure}[]	
\includegraphics[width=3.3in]{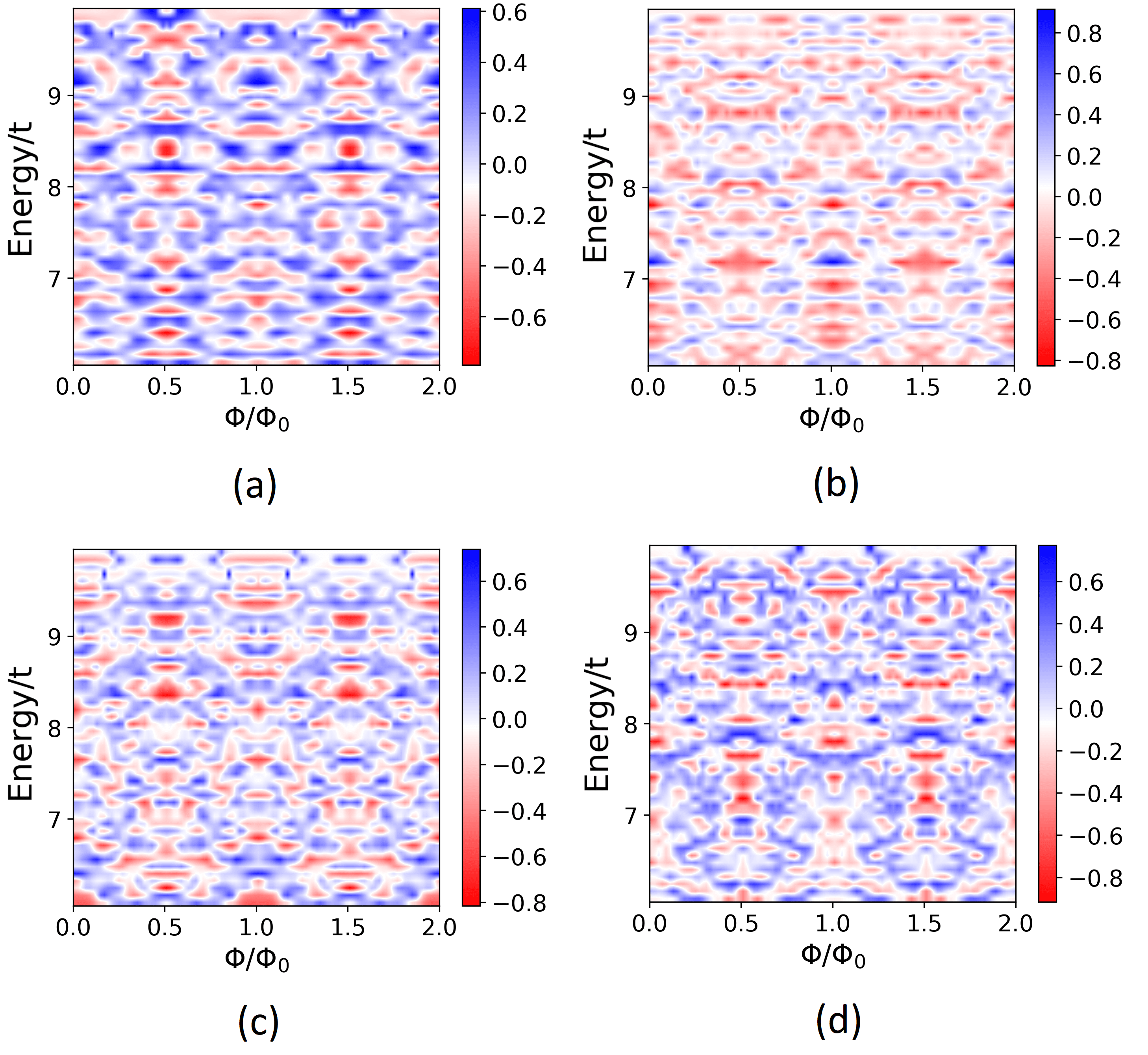}
\caption{$T(E,\Phi)-\langle T(E,\Phi)\rangle_{\Phi}$ for $W=1.5t$ and lengths of the disordered section equal to (a)$100a$, (b)$125a$, (c)$150a$ and (d)$175a$, in nanowires with strongly surface-confined electrons. For comparison, the strongly confined clean case is shown in Fig.~\ref{SR1}(a). With increasing length, AB oscillations ($(\Phi/\Phi_0)^{-1}=1$) weaken while AAS ($(\Phi/\Phi_0)^{-1}=2$) oscillations become significant. This leads to a transition point where AAS oscillations start dominating AB. The FFT spectrums shown in Fig.~\ref{T FFT AAS ALL} highlight the corresponding harmonic contents. }
\label{T AAS ALL} 
\end{figure}
\begin{figure}[]	
\includegraphics[width=3.5in]{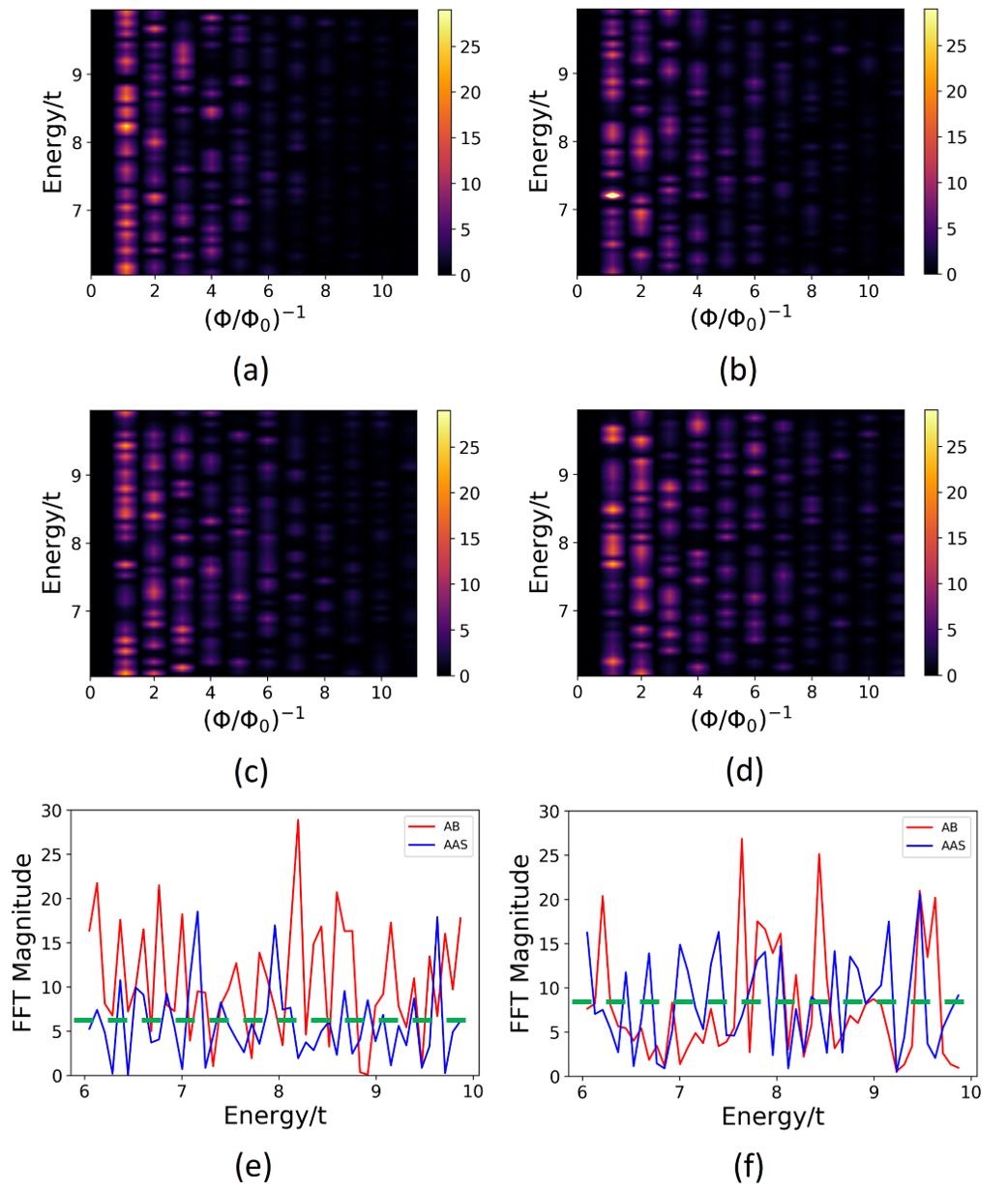}
\caption{FFT spectrum of the variation in transmission (Fig.~\ref{T AAS ALL}) from the mean for each energy, for $W=1.5t$ and lengths of the disordered section equal to (a)$100a$, (b)$125a$, (c)$150a$ and (d)$175a$, in nanowires with strongly surface-confined electrons. With increasing length, AB oscillations ($(\Phi/\Phi_0)^{-1}=1$) weaken while AAS ($(\Phi/\Phi_0)^{-1}=2$) oscillations become significant. This leads to a transition point where AAS oscillations start dominating AB. This is seen more clearly in panels (e) $100a$ and (f) $175a$, where the limiting cases for the AB and the AAS FFT are shown with the corresponding energy-averaged AAS magnitude shown in green (See Fig.~\ref{AB_AAS_comp}). Here (a) and (e) show the limit where the AB oscillations dominate, while (d) and (f) show the limit where the strength of the AAS oscillations become significant and comparable to the AB oscillations.}
\label{T FFT AAS ALL} 
\end{figure}
\begin{figure}[]	
\includegraphics[width=1.66in]{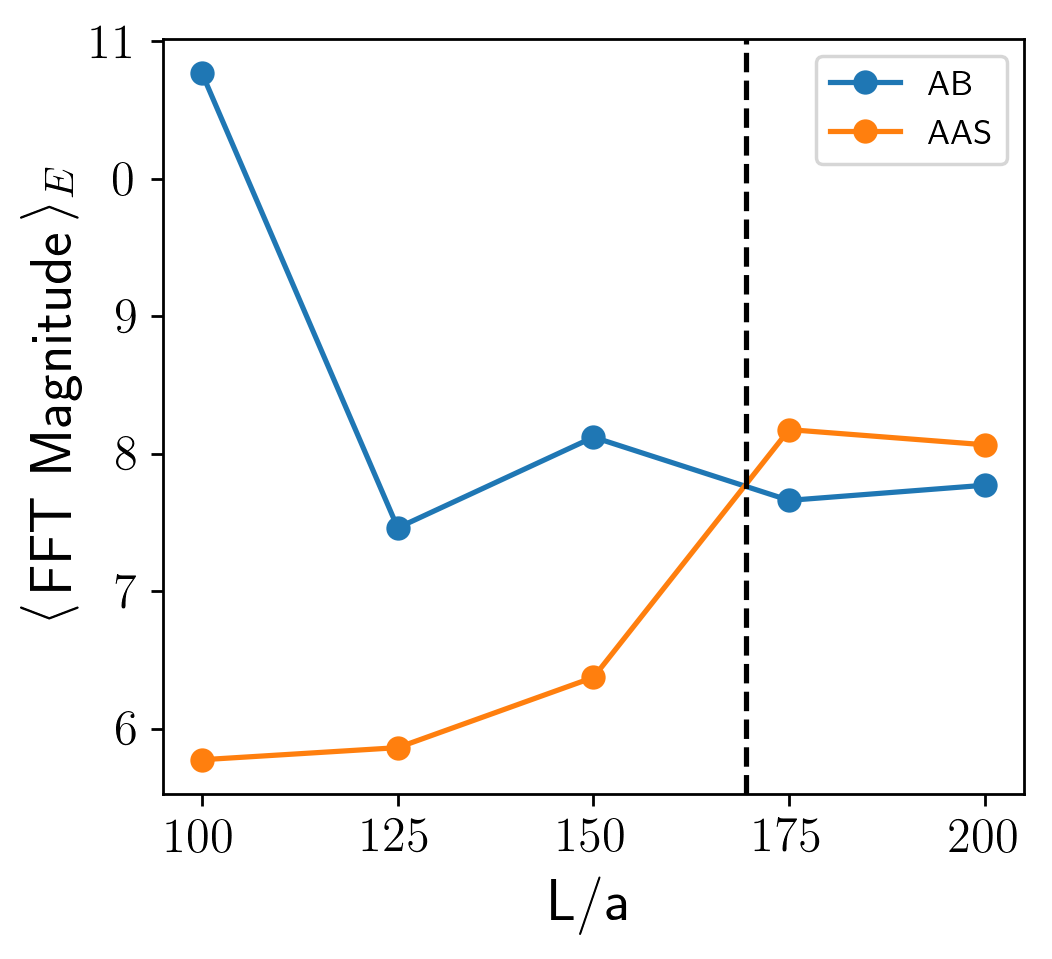}
\caption{FFT of the variation in transmission from the mean for each energy, averaged over energy, for $W=1.5t$, corresponding to the systems considered in Fig.~\ref{T FFT AAS ALL}. Note that for this disorder value, the nanowires are in the quantum diffusive/weakly localized regime for all energies. One can observe the transition point (shown by the dashed line) beyond which the AAS starts dominating over the AB oscillations. The increase in the magnitude of the AAS component is an effect of increasing phase coherence length with increasing nanowire length, as we are considering phase coherent transport here. Further, the residual AB component in long nanowires can be attributed to the leads, which are clean extensions of the device, and are subject to the same magnetic field.}
\label{AB_AAS_comp} 
\end{figure}

Coherent scattering cannot be captured by a self-energy as it is in general complex. The imaginary part of the self-energy will destroy the coherence~\cite{DATTA2000253,PhysRevB.75.081301}. Since the disorder is static, it is introduced using a random on-site uncorrelated disorder potential into the Hamiltonian, instead of using a self-energy~\cite{PhysRevB.60.16039}. The disorder potential is uniformly distributed in $[-W/2, W/2]$, where $W=\mathrm{constant}\times t$, and satisfies $\langle V(r)\rangle=0$ and $\langle V(r)V(r')\rangle=(W^2/12)\delta(r-r')$. Details regarding the implementation of the self-energy are explained in Appendix ~\ref{selfE}. For sufficiently long disordered nanowires in the quantum diffusive (phase coherent elastic scattering) regime (range of energies where the scattering length is smaller than the nanowire dimensions), AAS (h/2e) component survives, with a significant decrease in the AB (h/e) component due to `ensemble averaging'/phase-cancellation over the constituent rings/slices.

In Fig.~\ref{T AAS ALL} we show $T(E,\Phi)-\langle T(E,\Phi)\rangle_{\Phi}$ for lengths ranging from $100a-175a$. Fig.~\ref{T FFT AAS ALL} presents the corresponding FFT spectrums. Beginning with this section, we consider a nanowire with strongly surface-confined electrons for ease of simulation, using a much stronger transverse confining potential than the parabolic transverse potential used in Fig.~\ref{T FFT truepar circstep}. This explains the sharpness of the harmonics. In Fig.~\ref{T AAS ALL}(a) with $L=100a$ the effects of disorder are not significant. However, in Fig.~\ref{T AAS ALL}(d) with $L=175a$ disorder significantly alters the transmission spectrum, resulting in components with twice the AB frequency. Correspondingly, from Fig.~\ref{T FFT AAS ALL} with $L=100a$, it is observed that AB oscillations still dominate, even in the coherent scattering regime. A small degradation in AB amplitude is observed for $L=125a$. However, for $L=175a$, it is observed that in the quantum diffusive  and weakly localized regime, the amplitude of AAS oscillations just exceed that of the AB oscillations, signifying a transition point. The AB oscillation amplitude decreases as a longer nanowire ensures better averaging. This allows us to find a critical length, beyond which the AAS oscillations dominate. This is highlighted in Fig.~\ref{AB_AAS_comp}. Note that we observe a plateau forming for the AB component. This may be explained by the presence of the lead self energy in the device Green's function. The leads form a clean extension of the devive, with the same geometry, and is subject to the same magnetic field. Hence, the lead self-energy introduces a spatially in-homogeneous, flux dependent, non local quantity at the sites connected to the leads, unlike the analytical calculations performed for a closed system~\cite{JETP.35.11.588,RevModPhys.59.755}. This results in a non-vanishing contribution of the diffusons to the conductivity. Also, we observe a rise in the AAS component.  From~\eqref{coopcondcorr}, the magnitude of AAS$\sim\mathrm{K}_0\left(2\times2\pi R/l_\phi\right)$, which is a decreasing function of $\left(2\times2\pi R/l_\phi\right)$. Since we are considering phase coherent transport, we are inadvertently increasing the phase coherence length with increasing device length. This can roughly explain the rise, but one must be careful as it assumes complete averaging. From a different perspective, the diffusion probabilities are not restricted to the device alone. Since we are considering phase coherent transport in the device, the winding of the electronic paths are directly affected by the leads, as electrons may escape into the leads along their paths. Lastly, the non-monotonic variations in the plot are a consequence of the stochastic nature of the problem in the presence of uniform random disorder.

\begin{figure}[h!]	
\includegraphics[width=3.3in]{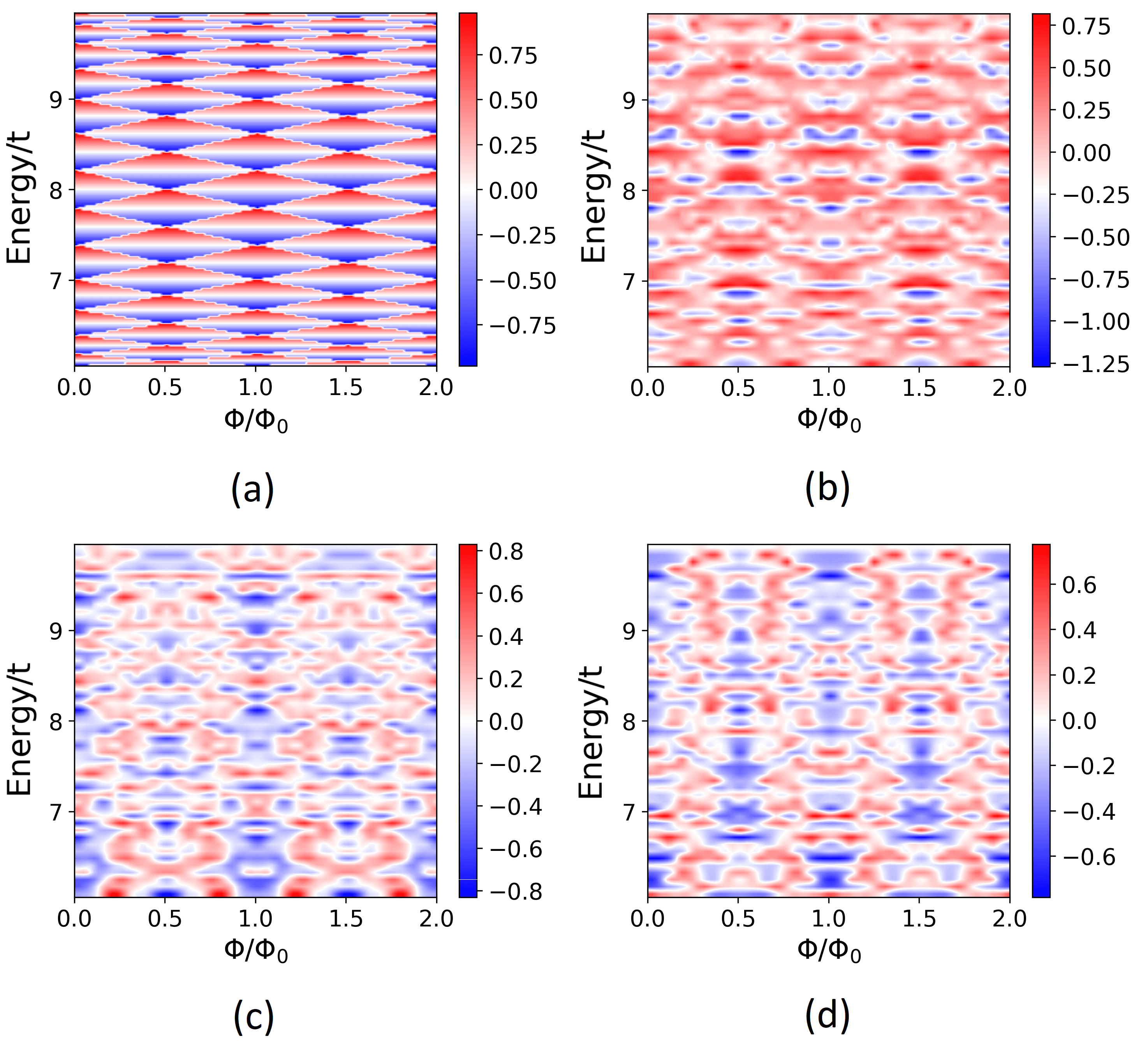}
\caption{$T(E,\Phi)-\langle T(E,\Phi)\rangle_{\Phi}$ for $L=100a$ and a range of disorder strengths (a)$W=0.0t$, (b)$W=1.0t$, (c)$W=1.6t$ and (d)$W=1.9t$. One can observe the transition from the AB dominated regime to the AAS dominated regime on increasing $W$. This is inferred by counting the repeating features with increasing flux at any given energy, which changes from being one in panel (a) to two in panel (d). Also, in (c) the AAS harmonic starts dominating from the center of the band $(E = 6t)$ while the AB harmonic is stronger near the band-edges (only the upper band-edge at $E=10t$ is shown here). This is explained by the variation of the scattering length over the band as shown in Fig.~\ref{Lscat} and the following corresponding discussion. Further, the AAS oscillations bear the same phase relation at all energies.}
\label{T AAS ALL 1} 
\end{figure}
\begin{figure}[h!]	
\includegraphics[width=1.7in]{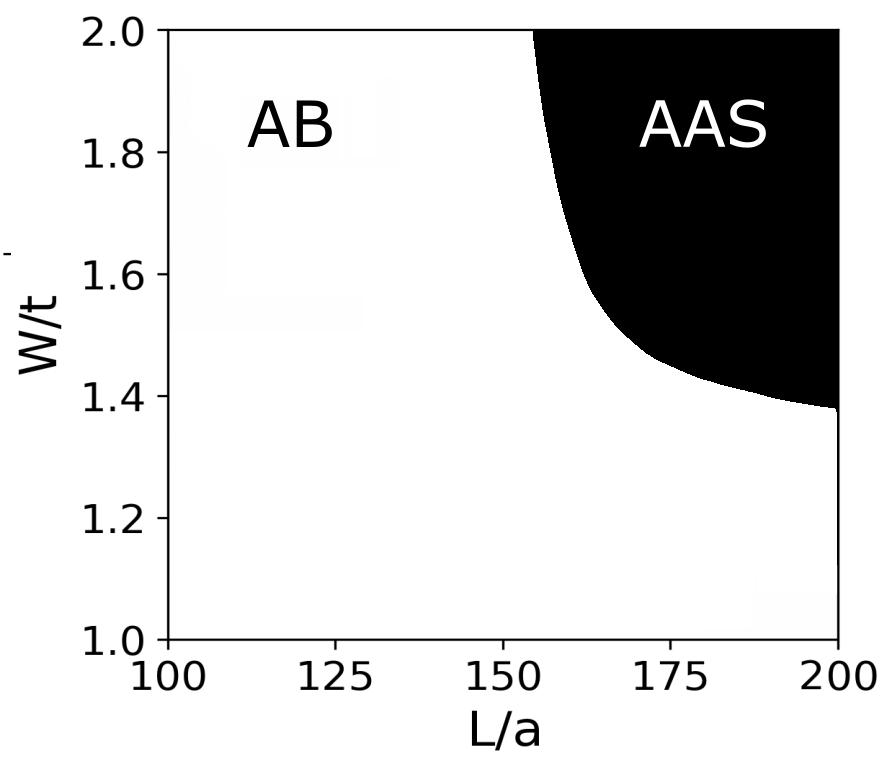}
\caption{The dominant harmonic contribution as a function of the disorder strength and device length (diameter $=10a$), as obtained from our numerical results. The boundary between the regions has been smoothly interpolated. The domain yielding a dominant AAS contribution is expected to increase on using a nanowire with a larger diameter due to a reduction in the scattering length.}
\label{T AAS ALL phase} 
\end{figure}

In Fig.~\ref{T AAS ALL 1} we observe the seamless transition from the AB dominated regime to the AAS dominated regime on increasing the disorder strength. Further, by gradually introducing disorder into the
system, the results progressively start to resemble experimentally observed features~\cite{PhysRevB.91.045422, EPLexp}. Note that while the AB oscillations at different energies are uncorrelated, the AAS oscillations are correlated~\cite{PhysRevLett.56.386}. This may be observed by observing the zero field phase of the harmonics. In Fig.~\ref{T AAS ALL phase} we show the numerically obtained dominant harmonic as a function of the length of the disordered section and the disorder strength, summarizing their roles. According to Fig.~\ref{T AAS ALL phase}, on increasing the length of the nanowire in Fig.~\ref{T AAS ALL 1} the characteristics of disorder should start appearing at smaller values of the disorder strength. Further, we find that a rather strong disorder potential is required for a dominant AAS contribution. However, the domain in $W$ and $L$ yielding a dominant AAS contribution is expected to increase on using nanowires with a larger diameter. The scattering length for a sub-band is given by,
\begin{align}
l_e&=\frac{\hbar v(E)}{2\pi}\frac{1}{a^2\frac{W^2}{12}\mathrm{DOS}(E)}\label{lscatt1},
\end{align}
where $v(E)$ is the group velocity and $\mathrm{DOS}(E)$ is the density of states of the considered sub-band at energy $E$. The net scattering legnth, after considering all the sub-bands, is given by a harmonic sum of the scattering lengths of each sub-band~\cite{nemec} (See Appendix~\ref{scattlsec} for details). Since the scattering length is inversely proportional to the density of states which in turn is proportional to the nanowire diameter, thicker nanowires will have a smaller scattering length. This is elaborated in the discussion following Fig.\ref{Lscat}. Note that while this analysis suggests that a larger nanowire shall support a significant AAS contribution, in reality, dephasing limits the system size over which phase coherence is retained. Consequently, the device size over which interference effects may be sustained is limited. The effects of dephasing are explored in detail in Sec.~\ref{dephasing}.

\begin{figure}[htb!]	
\includegraphics[width=3.3in]{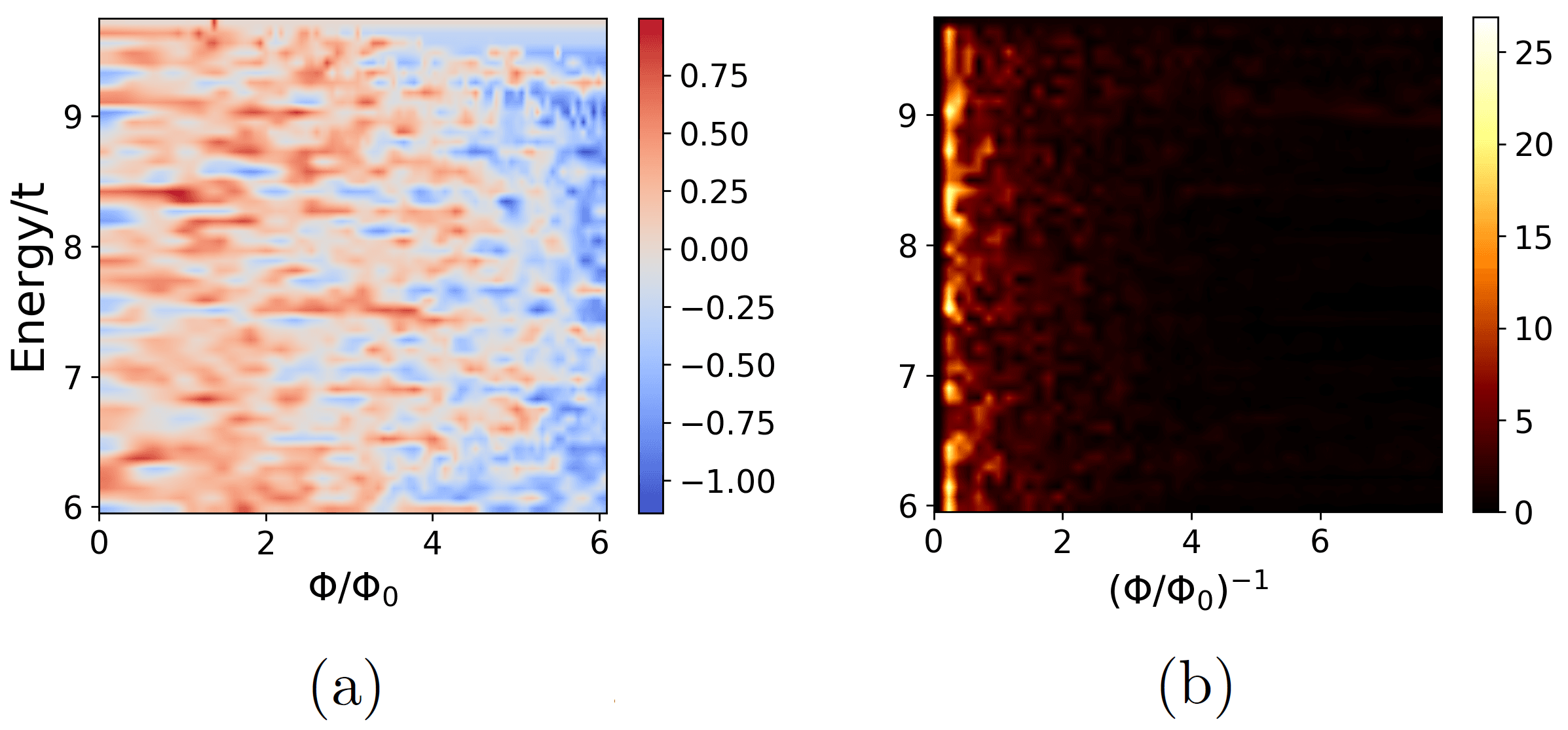}
\caption{Magnetic field perpendicular to axis ($B_{\perp}$) : (a) Transmission spectrum, and (b) its FFT, for a nanowire of length $125a$ with $W=1.5t$ and a strongly surface confined distribution. Note that $\Phi=B_{\perp}\pi R^2$. We observe aperiodic UCF, which vary much more slowly and randomly than the periodic AB oscillations.
}
\label{disorderBperp}
\end{figure}

Also, we observe that the FFT spectrum is dominated by the oscillation harmonics, with the UCF hardly contributing. A strongly surface confined distribution is not expected to contribute to UCF, as closed loops on the nanowire surface which do not traverse the circumference do not have a net flux through them ($\oint\mathbf{A}\cdot d\mathbf{l}=0$). Such features may be obtained from bulk transport with an axial magnetic field, or from both the bulk and the surface contributions in the case of non-axial magnetic field. In Fig~\ref{disorderBperp}, the results for a disordered nanowire with a magnetic field perpendicular to the axis is shown. In this case, magneto-conductance oscillations are dominated by loops confined on the surface, but not covering the circumference, showing only the UCF component. Such results have been experimentally realized\cite{Haas2016}.

\indent Back to the case of axial field, in the same transmission spectrum, two separate regimes can be observed. Near the center of the band, AB oscillations dominate, whereas on moving away from the enter, AAS oscillations dominates. This can be explained by finding the scattering length as a function of energy, as shown in Fig.~\ref{Lscat} (detailed in the Appendix~\ref{scattlsec}). Here, by the scattering length, we are referring to the mean distance between elastic scattering events. This is different from the transport mean free path, associated with back-scattering (describing the transmission in the diffusive limit\cite{doi:10.1063/1.3291120}), differing by an energy dependent relationship.
\begin{figure}[htb!]	
\includegraphics[width=3.25in]{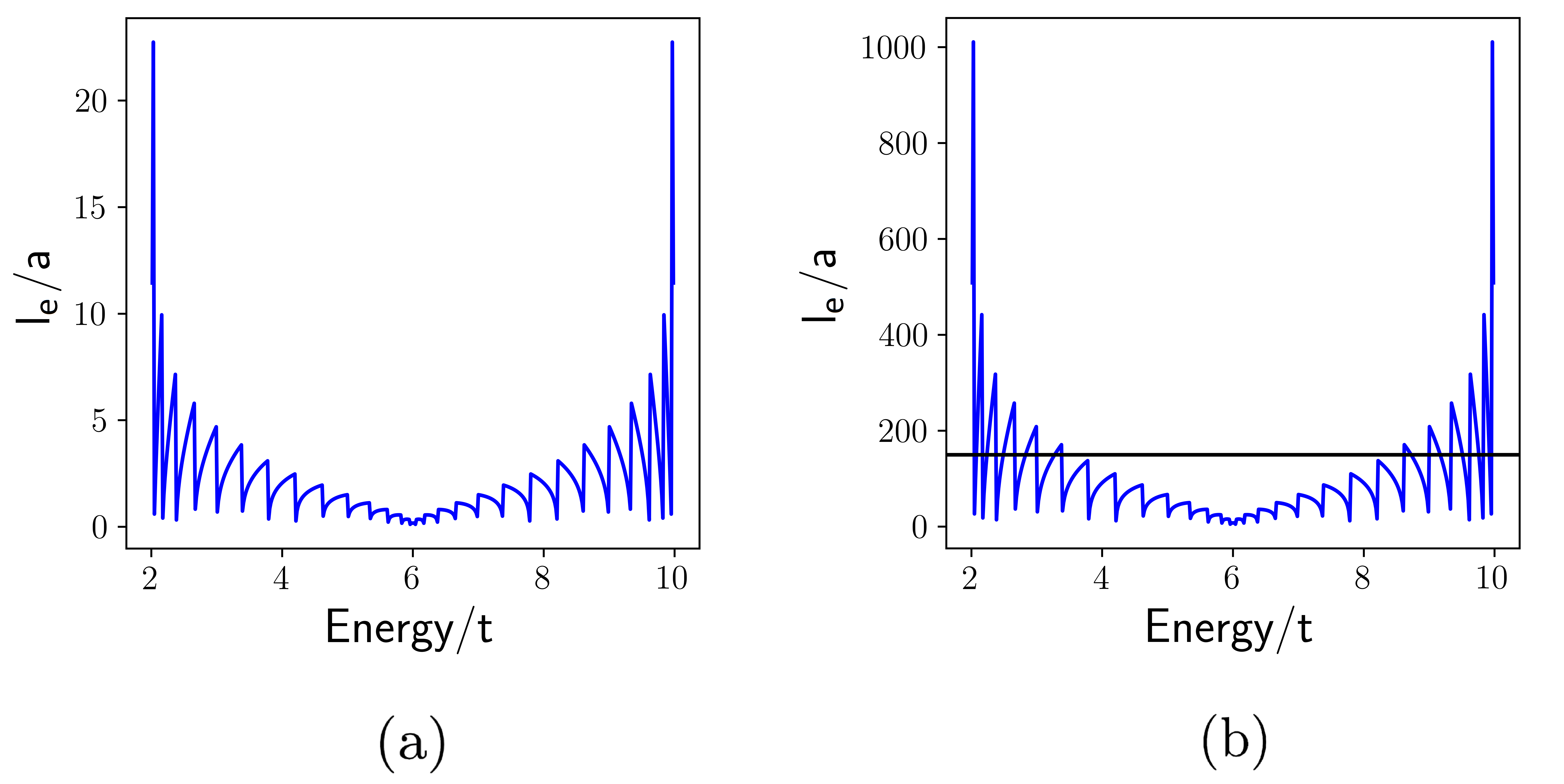}
\caption{Scattering length (in units of the lattice constant($a$)) as a function of energy in the band, for (a)$W=1.0t$, and (b)$W=0.15t$ (disorder potential $\in [-W/2,W/2]$). The band edges are located at $2t$ and $10t$. We see that transport is more diffusive in the middle of the band than near the band edges. In order to achieve a transition from the diffusive to the ballistic regime within the band, we must consider a much weaker scattering potential. In (b), we have marked the length $L=150a$, which is the length of the nanowire used in Fig.~\ref{Taas_regimetrans}. Note that the dips in the scattering length correspond to the Van Hove singularities, where the associated
high density of states enhances scattering.}
\label{Lscat}
\end{figure}
\begin{figure}[htb!]	
\includegraphics[width=3.45in]{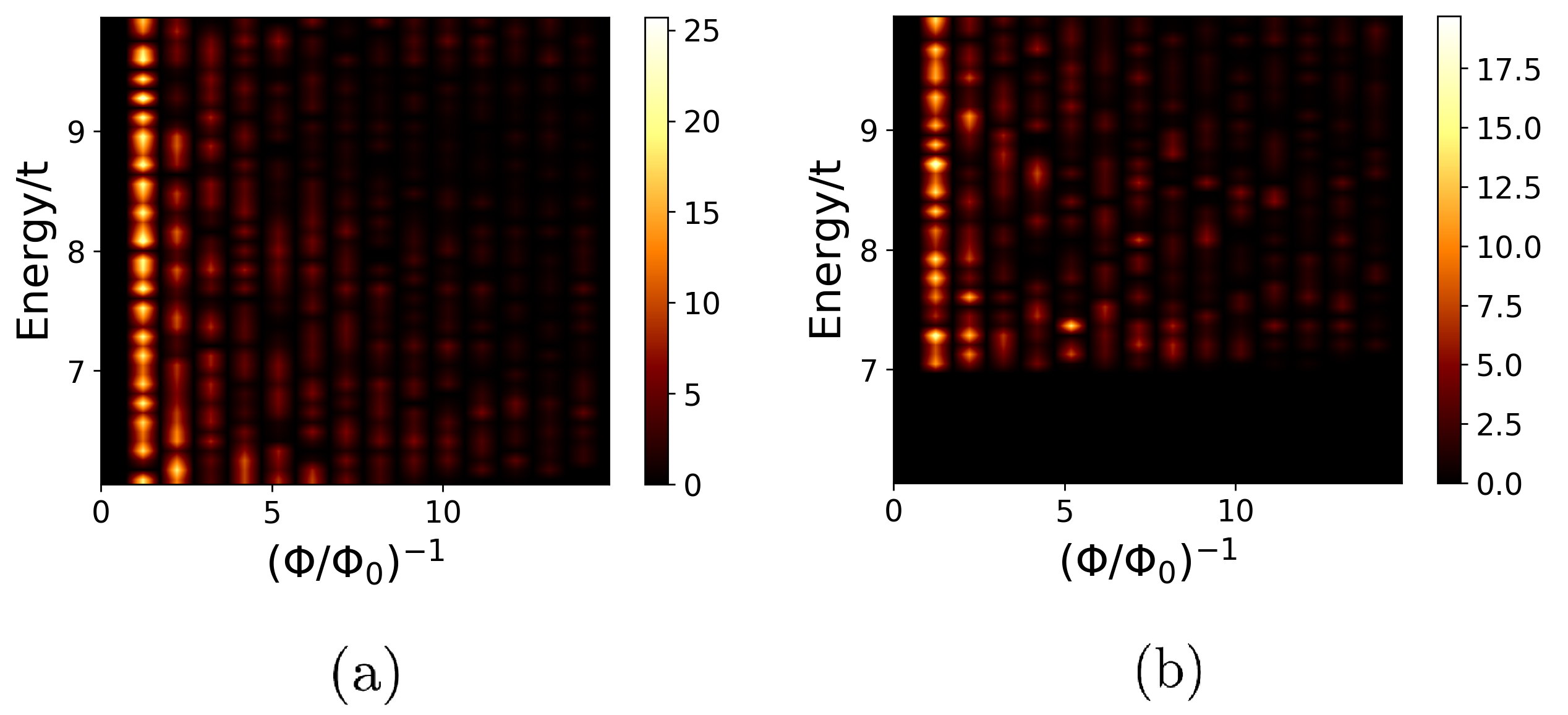}
\caption{FFT spectrum of the transmission: (a) For a nanowire of length $150a$, with disorder potential given by $W=0.15t$. A significant AAS component is observed in the diffusive regime near the center of the band ($E=6t$), which subsides as we move higher in energy to the ballistic regime near the band-edge ($E=10t$). Only in the range of energies where the scattering length is smaller than the nanowire length (marked in Fig.~\ref{Lscat}), which is roughly $E\in[3t,9t]$, we observe a significant AAS component. Outside this range ($E>9t$ in this figure), the AAS component is much smaller than the AB component.  (b) For the same nanowire as in (a), but with the on-site energies ($E=5t$ and $E=7t$) alternating on neighboring sites. Apart from the bandgap at the center of the band, the AAS oscillations dominate in two separate regions around $E=7t$ and $E=9t$.}
\label{Taas_regimetrans}
\end{figure}

From Fig.~\ref{Lscat}, we see that the scattering length is smallest near the band center and increases as we move away from the center. This implies that the band center experiences a more diffusive environment than the band edge. This is a consequence of the effective two-dimensional nature of the problem due to strong surface confinement, leading to a growing mobility edge appearing from the band center. A shorter scattering length near the band-center leads to a dominant AAS contribution in the center of the band and not near the band-edges, where the AB component strengthens. This is clearly seen clearly in Fig.~\ref{T AAS ALL 1}. Since experiments typically probe the conduction band-edge (lower edge), observing the AAS harmonic becomes harder. On considering a weaker scattering potential, we can observe the diffusive to ballistic transition in the transmission spectrum, as seen in Fig.~\ref{Taas_regimetrans}(a). Further, the strength of the AAS harmonic is found to be related to the the relative magnitude of the nanowire length and the scattering length. Also, since~\eqref{lscatt1} implies that the scattering length decreases with increasing density of states (detailed in Appendix~\ref{scattlsec}), which in turn increases with increasing nanowire diameter; a nanowire with a larger diameter is expected to show stronger AAS oscillations. In Fig.~\ref{Taas_regimetrans}(b), where we consider the same nanowire as in Fig.~\ref{Taas_regimetrans}(a), but with alternating on-site energies on neighboring sites, we observe a different behavior as a result of the change in the density of states. The AAS oscillations are significant in two regions, around $E=7t$ and $E=9t$, with its strength displaying a non-monotonic behavior unlike Fig.~\ref{Taas_regimetrans}(a). This highlights the effect of dispersion of surface states in governing the strength of the harmonics with energy. These observations may be used as a guideline to extract the elastic scattering length from experiments in suitable situations, by varying the gate voltage and studying the oscillatory components. 

Assuming ideal surface confinement, this structure of the transport regime seen across the band is a function of the surface dispersion, which enters the scattering length via its dependence on the density of states and the group velocity. Hence a different lattice would have a different structure.

Note that, throughout the work, we do not look at the disorder induced band-tail extensions as the leads are infinitely long and clean extensions of the device, which do not support such states.
\subsubsection*{\label{wl}\textbf{Oscillatory/non-oscillatory weak localization}}
\begin{figure}[htb!]	
\includegraphics[width=3.44in]{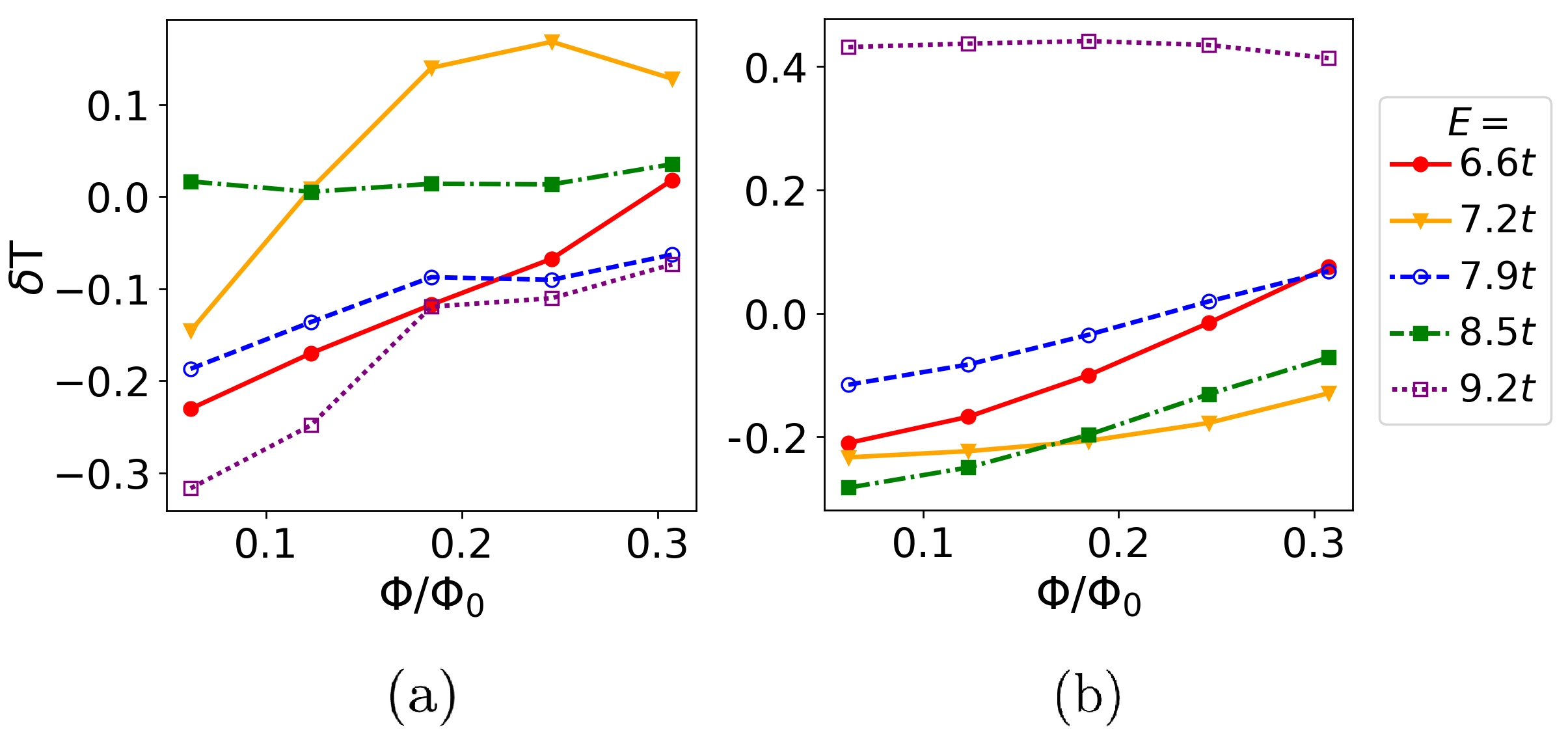}
\caption{$\delta T(E,\Phi)=T(E,\Phi)-\langle T(E,\Phi) \rangle_\Phi$ : (a) Parallel field : Nanowire of length $175a$, $W=1.5t$, and strong surface confinement. We have introduced a finite temperature by averaging over an energy interval$ =0.31t$. (b) Perpendicular field : Nanowire of length $125a$, $W=1.5t$, and strong surface confinement, corresponding to the system in Fig.~\ref{disorderBperp}, with no energy averaging. As expected, we observe that the transmission initially rises with increasing flux, as the phase relationships between the disorder induced localized states are destroyed. Note that the magnitude of the weak localization correction is $\approx 0.1-0.3$, which can get easily subdued by the oscillatory components. Also, on extending panel (a) to higher flux values, we recover oscillatory behavior which eventually start dominating. But on extending (b) we get the same transmission values as shown in Fig.~\ref{disorderBperp} as it's the same system.}
\label{weaklocalT}
\end{figure}

In a nanowire with a surface confined distribution, the oscillation harmonics introduce transmission variations of magnitude $\approx 1$. Non-oscillatory weak localization (WL) corrections, which give rise to a decreasing resistance with increasing magnetic field, introduce much smaller variations. In order to observe them separately without being subdued by the large oscillatory AB component and UCF, we have to reduce the oscillations. This may be achieved by doing the following:
\begin{enumerate}[{(1)}]
\item Using long disordered nanowires to suppress the AB component.
\item Along with point (1) averaging the transmission spectrum over a range of energies, as shown in Fig.~\ref{weaklocalT} to kill random fluctuations (UCF). This also helps to kill the AB harmonic as the oscillations at differennt energies are uncorrelated. Such averaging effects maybe present in experiments due to non-zero temperatures, or finite applied bias. Finite energy averaging has been studied earlier in similar contexts~\cite{PhysRevB.53.R1693}. Note that the origin of AAS oscillations is same as the conventional WL and in the low field regime only the latter manifests itself.
\item Using an electronic distribution which naturally results in diminished oscillatory components. This may be brought about by weaker surface confinement, as mentioned in Sec.~\ref{clean}. 
\item Using a perpendicular magnetic field, which would suppress all harmonics.
\end{enumerate}
In Fig.~\ref{weaklocalT} we show the non-oscillatory weak localization corrections after suppressing the oscillatory ones. At few values of energy, we still observe an initially decreasing transmission, which may be attributed to a large UCF component. With energy averaging, the transmission rises at all energies, for small values of flux. Further, as mentioned above the origin of AAS oscillation are same as the pure weak-localization corrections. Therefore the for small flux values, the transmission traces in Fig.~\ref{T AAS ALL 1} (d) as well as Fig.~\ref{SR2}(a) rise as well.

\subsection{\label{sr}SURFACE ROUGHNESS}
\begin{figure}[htb!]	
\includegraphics[width=3.35in]{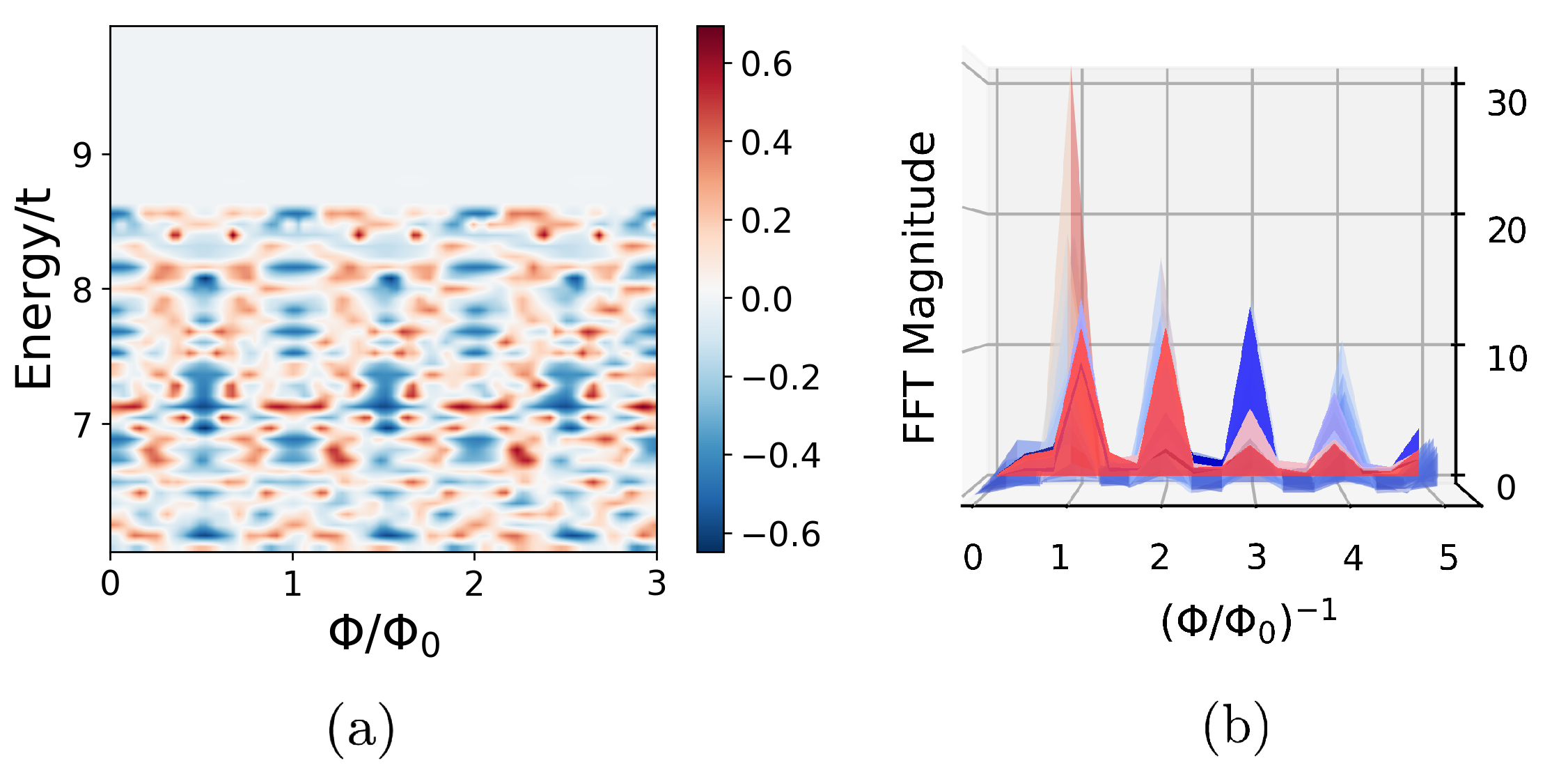}
\caption{Surface roughness without on-site Anderson disorder: (a)$\delta T(E,\Phi)$, and (b) its FFT, of a clean nanowire of length $L=100a$, with surface roughness described by $\zeta=-2$ and $\Delta_\mathrm{R} = 0.4\mathrm{R}_0$. We observe a degradation of the AB component, accompanied by the dominance of the AAS component in general introduced primarily by the hopping disorder (except at $E=7.1t$, which shows up as an odd AB peak in the FFT spectrum, at that energy). The AAS contribution is clearly visible in Fig.~\ref{SR2}(a). Note that, we do not observe magnetic depopulation, as seen in Fig~\ref{SR1}. Further, as explained in the text, the skewed hopping disorder reduces the bandwidth of transmission.}
\label{SR2}
\end{figure}

\begin{figure}[htb!]	
\includegraphics[width=3.4in]{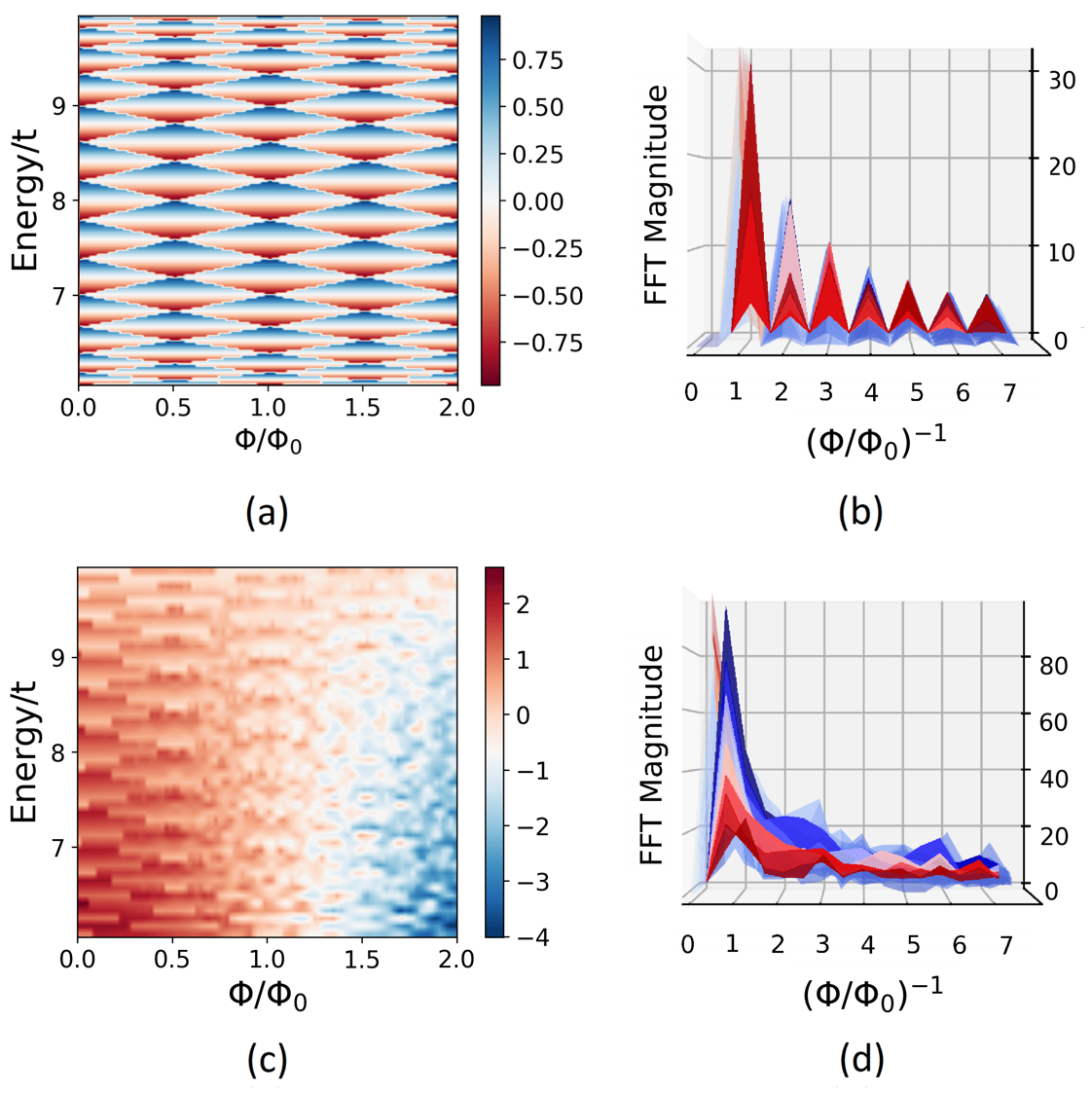}
\caption{Surface roughness without on-site Anderson disorder: (a) Variation of the transmission from the mean value for each energy $\delta T(E,\Phi)$, and (b) its FFT, without SR, in a nanowire with strong surface confinement for comparison. The 'discretized' diamonds are due to the presence of a finite number of states. In (c), we see $\delta T(E,\Phi)$, and (d) its FFT, of a clean nanowire of length $L=100a$, but with surface roughness, described by $\zeta=0$ and $\Delta_\mathrm{A} = 0.4\mathrm{A}_0$. Note that, in panels (b) and (d), the energy axis goes into the page. Here $\mathrm{E}=6t$ and $\mathrm{E}=10t$ are the band-center and the upper band-edge respectively. A degradation of the sharp features is observed in the transmission spectra as a result of SR, which is manifested as a smoother decay of the corresponding FFT spectrum. The strong low frequency peaks ($(\Phi/\Phi_0)^{-1}<1$) correspond to the slow drop in transmission due to magnetic depopulation. The magnitudes of the harmonics have been found to be same. Further, at higher values of flux, the AB oscillation diamonds have degraded.}
\label{SR1}
\end{figure}

We now proceed to detail the effects of surface roughness/random corrugations. We model the surface roughness by a radius $R(\phi,z)$, which varies randomly as a function of the azimuth angle and the axial distance. This induces random variations in the hopping parameters and the flux across each cross-sectional disc along the nanowire axis. The variations in the radius are described by an uncorrelated noise, with $R(\phi,z)-R_0\in [-\Delta_R/2,\Delta_R/2]$, such that $\langle R(\phi,z)\rangle=R_0$ ($R_0$ is the radius without surface roughness), and $\langle R(\phi,z)R(\phi',z')\rangle=(\Delta_\mathrm{R}^2/12)\delta(\phi-\phi')\delta(z-z')$. Note that we do not consider Anderson disorder while studying surface roughness, resulting in a diffusive regime due to random hoppings instead of random on-site potentials. Now, we adopt a generalized Harrison~\cite{harrison} (power-law) scaling of the hopping parameters with respect to the corresponding bond distances, which gives us,
\begin{align}
\tilde{t}_{\mathbf{x},\mathbf{x}'}&=t_{\mathbf{x},\mathbf{x}'}\left(\frac{\sqrt{a^2+(r(\mathbf{x})-r(\mathbf{x}'))^2}}{a}\right)^\zeta, \label{harrison}
\end{align}
where $\mathbf{x}\equiv \left(\phi,z\right)$, and $a$ is the bond distance in the absence of surface roughness. $\zeta<0$ represents the sensitivity of the hopping parameter to variations in the bond distance. We introduce two cases cases which we discuss subsequently: First, $\zeta=-2$ along with minor realistic surface roughness (specified in terms for radial variations as the corresponding areal variations are negligible) in Fig.~\ref{SR2}; second $\zeta=0$ along with severe roughness (specified in terms of transverse areal variations) in Fig.~\ref{SR1}.

In Fig.~\ref{SR2}, we implement a realistic surface roughness using $\Delta_\mathrm{R} = 0.4\mathrm{R}_0$ and $\zeta=-2$, which induce minor variations in the cross-sectional area with standard deviation $\sigma_\mathrm{A}=-0.04\mathrm{A}_0$, and variations in the hopping given by $\sigma_\mathrm{\tilde{t}}= 0.23\mathrm{t}$. 
When the variation in the flux experienced by different planes is of the order of the flux quantum $(\Phi_0=h/e)$, it should destroy the phase relationships and consequently the flux periodic oscillations. However, for all practical values of $\Delta_{\mathrm{R}}$, the corresponding variation in the cross-sectional surface area and hence the flux penetrating each cross-sectional disc is too small to induce significant flux variations. Therefore, the effects of surface roughness are dominated by random hopping parameters. In this case (shown in Fig.\ref{SR2}), we observe a degradation of the AB component, accompanied by a significant contribution from the AAS component. Moreover, while on-site Anderson disorder and surface roughness both lead to the emergence of AAS oscillations, only the latter decreases the bandwidth (difference of upper and lower band-edges) of transmission which decreases on the mean value of the hopping parameters. The reduction in the bandwidth is a consequence of the power-law dependence of the hopping parameters on the bond distances (see~\eqref{harrison}) which reduces its mean value thereby skewing its distribution.
 
Now, in Fig.~\ref{SR1}(c) and (d), we implement strong roughness with the cross-sectional area $\mathrm{A}(\mathrm{z})\in [-\Delta_A/2,\Delta_A/2]$ (where $\Delta_A=0.4\mathrm{A}_0$) with a uniform distribution. It doesn't suppress the harmonics, but rather causes a large drop in the transmission induced by magnetic depopulation. The flux penetrating each cross-sectional disc along the nanowire axis is given by, $\tilde{\Phi}(\mathrm{z}) = \Phi(\mathrm{A}(\mathrm{z})/\mathrm{A}_0)$, where $\Phi$ is the flux in the absence of surface roughness, and $\mathrm{A}(\mathrm{z})$ is the cross-sectional area. Note that, the same variation in cross-sectional area induces larger variations in the flux, at higher values of flux as $\Delta_\Phi=B\Delta_A=\langle\Phi\rangle\Delta_A/\langle A\rangle$. The effects of this can be observed in Fig.~\ref{SR1}(c) from the stronger degradation of the AB component at higher values of flux, than at lower values. Also note that we consider perfect surface confinement. However, for large values of $\Delta_A$ as the electronic distribution is subjected to a large variation in the magnetic flux, the system may effectively be considered as possessing a radially smeared/spread-out electronic distribution (weaker surface confinement). From the discussion following Fig.~\ref{T FFT no pot circstep}, the magnetic depopulation is justified as the energy bands rise quickly with the flux on decreasing the surface confinement.

In conclusion, surface roughness can have varying and contrasting effects with the observed behavior being dependent on the material and the sample under consideration.. For considerable and realistic values of $\zeta$, the effect of surface roughness is dominated by the random hopping parameters, which results in the dominance of the AAS component. For values of $\zeta$ close to zero, we observe a magnetic depopulation. Further, the degradation of the AB component occurs only at higher values of flux while the AAS component never dominates. Also, large variations in the nanowire cross-sectional area may produce flux variations of the order of the flux quantum when the nanowire is subjected to strong magnetic fields, destroying the oscillatory magneto-conductance features. 

\subsection{DEPHASING}
\label{dephasing}
Magneto-conductance oscillations arise from the phase picked up over closed loops. For a particular harmonic to survive, the phase coherence length should be greater than the corresponding constituent path lengths. In~\eqref{coopcondcorr}, the amplitude of the harmonics is given by the Macdonald function $(\mathrm{K}_0(z=(n2\pi R/l_\phi)))$, which exponentially decays $(\sim e^{-z}/\sqrt{z})$ to zero for $z\rightarrow \infty$. This suggests that when the $l_\phi<2\pi n R$, then the amplitude of the corresponding oscillatory component with period $h/(2ne)$ exponentially vanishes. To clearly see its effect, phase relaxation (rather randomization) has been included, which should be able to diminish/eliminate the oscillations.

The dephasing is implemented in the self-consistent Born approximation (or the non crossing approximation) by a phenomenological dephasing model~\cite{PhysRevB.75.081301}$^{,}$~\cite{PhysRevB.85.155414}$^{,}$~\cite{crestipreldep}, to emulate electron-electron and electron-phonon interactions.

Beginning with the current in lead $x$, 
\begin{align}
I_x=\frac{2e}{\hbar}\int\frac{dE}{2\pi}\left[\Sigma_x^<(E)G^>(E)-\Sigma_x^>(E)G^<(E)\right], \label{curr1}
\end{align}
the kinetic equation for the lesser Green's function, with the lead and elastic interaction self-energies given by $\Sigma_{C}$ and $\Sigma_{S}$, respectively,
\begin{align}
&G^<(E)=G^R(E)\Sigma_{C}^{<}(E)G^A(E)+G^R(E)\Sigma_{S}^{<}(E)G^A(E), \label{dephase_selfE2}
\end{align} 
reduces the current to,
\begin{align}
I&=I_{\mathrm{coh}}+I_{\mathrm{incoh}},\\
I_{\mathrm{coh/incoh}}&=\frac{2e}{\hbar}\int\frac{dE}{2\pi}\Big[\Sigma_x^<(E)\left(G^R(E)\Sigma_{C/S}^{>}(E)G^A(E)\right)\nonumber\\
&-\Sigma_x^>(E)\left(G^R(E)\Sigma_{C/S}^{<}(E)G^A(E)\right)\Big].
\end{align}
Now, the general dephasing self-energy is given by,
\begin{align}
\left[\Sigma^R_S(E)\right]&= \tilde{D}\left[G^R(E)\right], \label{dephase_selfE1}\\
\left[\Sigma_S^{</>}(E)\right]&= \tilde{D}\left[G^{</>}(E)\right], \label{dephase_selfE3}\\
G^R(E)&=[(E+i\eta)I-H-\Sigma^R_C(E)-\Sigma^R_S(E)], \label{dephase_selfE0}
\end{align}
where $\tilde{D}$ is an operator, whose form depends on the dephasing scheme and strength. Now \eqref{dephase_selfE1}, \eqref{dephase_selfE3}, \eqref{dephase_selfE0} and \eqref{dephase_selfE2} may be solved numerically to get the current. However, further reduction gives us the following relations for the transmissions,
\begin{align}
I&=\int\frac{dE}{2\pi}\left(T_{\mathrm{coh}}(E)+T_{\mathrm{incoh}}(E)\right),\label{curr}\\
T_{\mathrm{coh}}(E)&=\mathbf{Tr}[\Gamma_L G^R(E)\Gamma_R G^A(E)],\\
T_{\mathrm{incoh}}(E)&=\mathbf{Tr}[\Gamma_L G^R(E)K(E) G^A(E)],
\end{align}
where $K(E)$ is obtained self-consistently by,
\begin{align}
K(E)&=\tilde{D}\left[G^R(E)\left(\Gamma^R(E)+K(E)\right)G^A(E)\right].\label{genK}
\end{align}
For the momentum relaxing scheme, the scattering self energy is local in its action and is therefore diagonal in its real space matrix representation. Therefore, $\tilde{D}[L]_{i,j}=D[L]_{i,j}\delta_{i,j}$, for a matrix $[L]_{i,j}$, with $D$ being the dephasing strength. This reduces \eqref{dephase_selfE3} and \eqref{genK} to,
\begin{align}
\left[\Sigma_S(E)\right]_{i,j}^<&=\sum_{k}M_{(j,k)}\left[G^{R}\Sigma_C^< G^{A}\right]_{k,k}\delta_{i,j},\\ 
\left[M\right]_{i,j} &= \frac{D}{I-D\mathopen|\left[G^R(E)\right]_{i,j}\mathclose|^2}\delta_{i,j},\\
K&=\sum_{k}\left[M\right]_{j,k}[G^{R}\Gamma_2 G^{A}]_{k,k}.
\end{align}
The scattering strength is given by $D\sim 1/\tau$, which is a measure of the correlation of the dephasing scattering potential $U(r)$, $D(E,r,r')=D\delta(r-r') \propto \braket{U(r)|U(r')}$.
\indent In semiconductor nanowires at low temperatures, low energy dephasing scattering is dominated by elastic electron-electron interactions\cite{0953-8984-14-18-201} and electron-phonon interactions~\cite{PhysRevLett.104.206803,0953-8984-14-18-201}. Also, acoustic phonon scattering is nearly elastic and randomizes the momentum of the electronic distribution. In that case, $D=\zeta^2k_BT/(\rho ^2)$, where $\zeta$ is the deformation potential, $\rho$ is the density, and $v$ is the longitudinal sound velocity. This permits us to consider these processes together in the phenomenological dephasing model, once the total scattering rate is accounted for $(1/\tau = 1/\tau_{e-e}+1/\tau_{e-ph})$. We have not used B\"uttiker probes, as they are phenomenological and appropriate for inelastic scattering, such as longitudinal electron-phonon (e-ph) interactions, which have been neglected in this study. 
\begin{figure}[htb!]	
\includegraphics[width=3.5in]{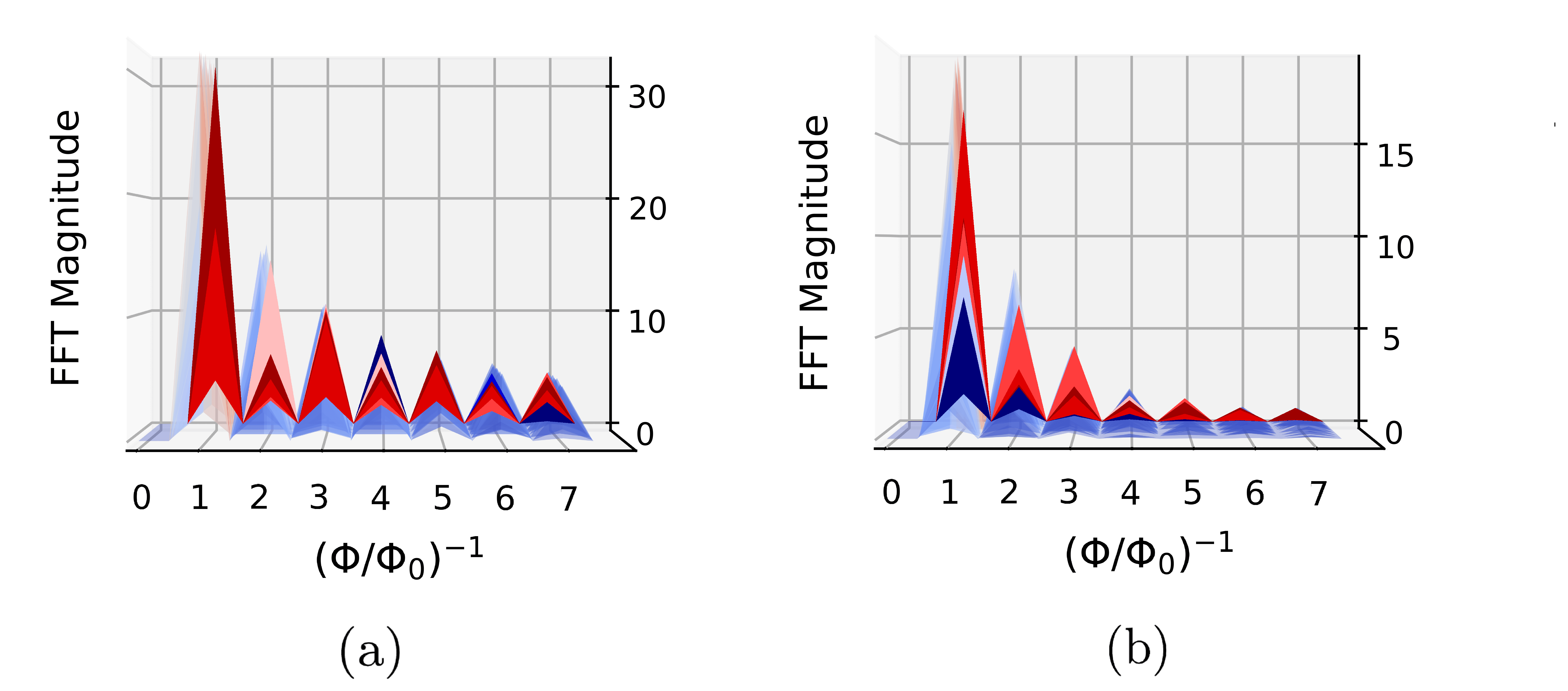}
\caption{Dephasing - Clean case : FFT spectrum of the variation in transmission from the mean value for each energy $\delta T(E,\Phi)$, (a)without dephasing and (b)with momentum relaxing dephasing in a clean nanowire of length $\mathrm{L}=50a$ with strong surface confinement. The energy axis goes into the page. The difference between (b) and (a) can be observed by comparing the decreased magnitudes of the Fourier components in (b).}
\label{clean dephase}
\end{figure}
\begin{figure}[htb!]	
\includegraphics[width=3.4in]{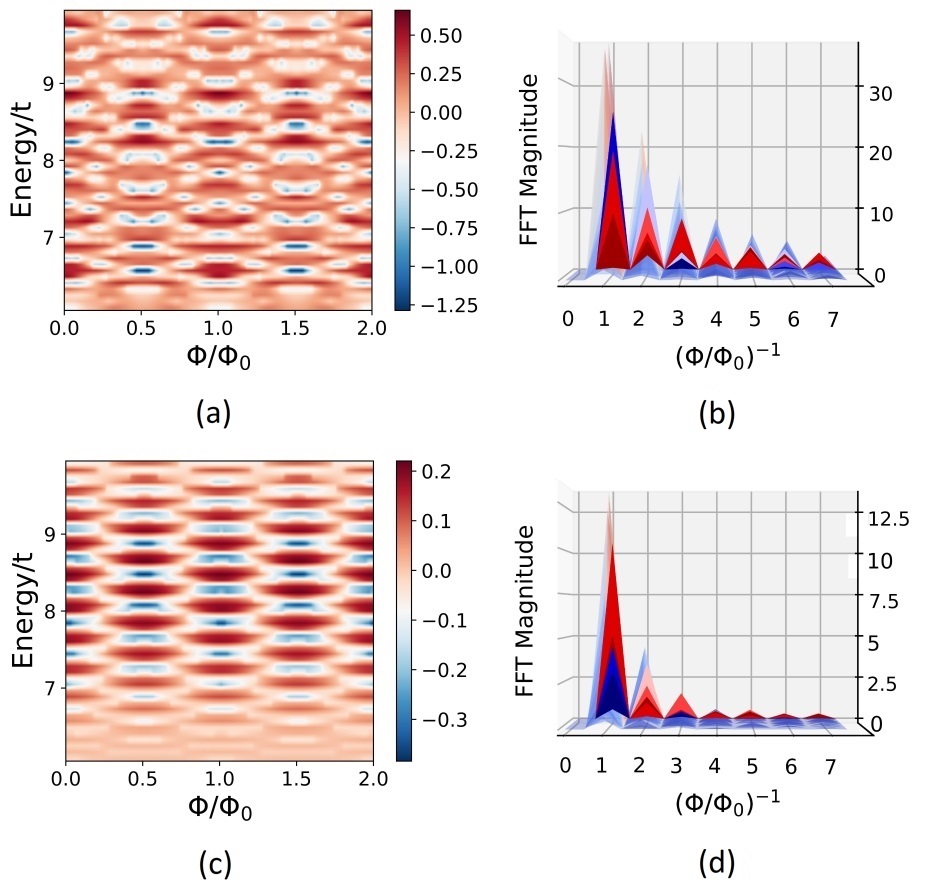}
\caption{Dephasing - Disordered case ($W=0.75t$) : (a) Variation of the transmission from the mean value for each energy $\delta T(E,\Phi)$, and (b) its FFT, without dephasing, in a disordered nanowire of length $\mathrm{L}=50a$ with strong surface confinement. In (c), we see $\delta T(E,\Phi)$, and (d) its FFT, of the same nanowire, but with momentum relaxing dephasing ($\mathrm{D}=0.4^2$). The energy axis goes into the page. A degradation in the amplitudes of the harmonics is observed in the FFT spectrum (from (b) to (d)). Further, the degradation for the AB peak is smaller than the higher harmonics, increasing its dominance. Consequently, the distinct diamond shaped structures representing the AB oscillations become the dominant feature in the transmission spectrum. Here $\mathrm{E}=6t$ and $\mathrm{E}=10t$ are the band-center and the upper band-edge respectively.}
\label{disprelincohdephase}
\end{figure}

Now, as the magnitude of dephasing is gradually increased, the phase coherence length should decrease, and fall behind the required length to sustain each winding number. This effect should be observable in the FFT spectrum as a systematic degradation of the oscillatory part, with the highest harmonics vanishing one by one on increasing the dephasing potential. In the case of clean nanowire, it is seen from Fig.~\ref{clean dephase} that the magnitude of the AB oscillations die down. Further, in the case of disordered wires, we observe a degradation in all the harmonics, with the higher harmonics degrading much faster than the AB harmonic. This leaves us with a relatively dominant AB contribution. This can be noticed by comparing Figs.~\ref{disprelincohdephase}(c) and (d). Also, at the edges of the steps in the transmission, we encounter Van Hove singularities in the density of states, which increase the scattering rate. This results in smoothed out steps.

We have until now, considered a dephasing rate which is constant with respect to energy. In reality, scattering rates depend on energy\cite{Singha2017}, being typically of the form $\tau(E)\propto E^r$. When an energy dependent scattering rate is taken into account in the local dephasing model, the degradation of the harmonics becomes energy dependent.

Note that the effect of dephasing is very different compared to the effect of surface roughness. While dephasing kills the higher harmonics one by one, creating a relative dominance of the AB component, surface roughness may lead to the relative dominance of the AAS component, similar to the case with local potential disorder considered in Sec~\ref{dis}. This is evident from Figs.~\ref{SR2} and~\ref{disprelincohdephase}. This observation may serve as a guideline to pinpoint the source of features observed in experiments.

\subsection{DISORDERED AND INCOHERENT NANOWIRES WITH WEAK SURFACE CONFINEMENT}
\begin{figure}[htb!]	
\includegraphics[width=3.5in]{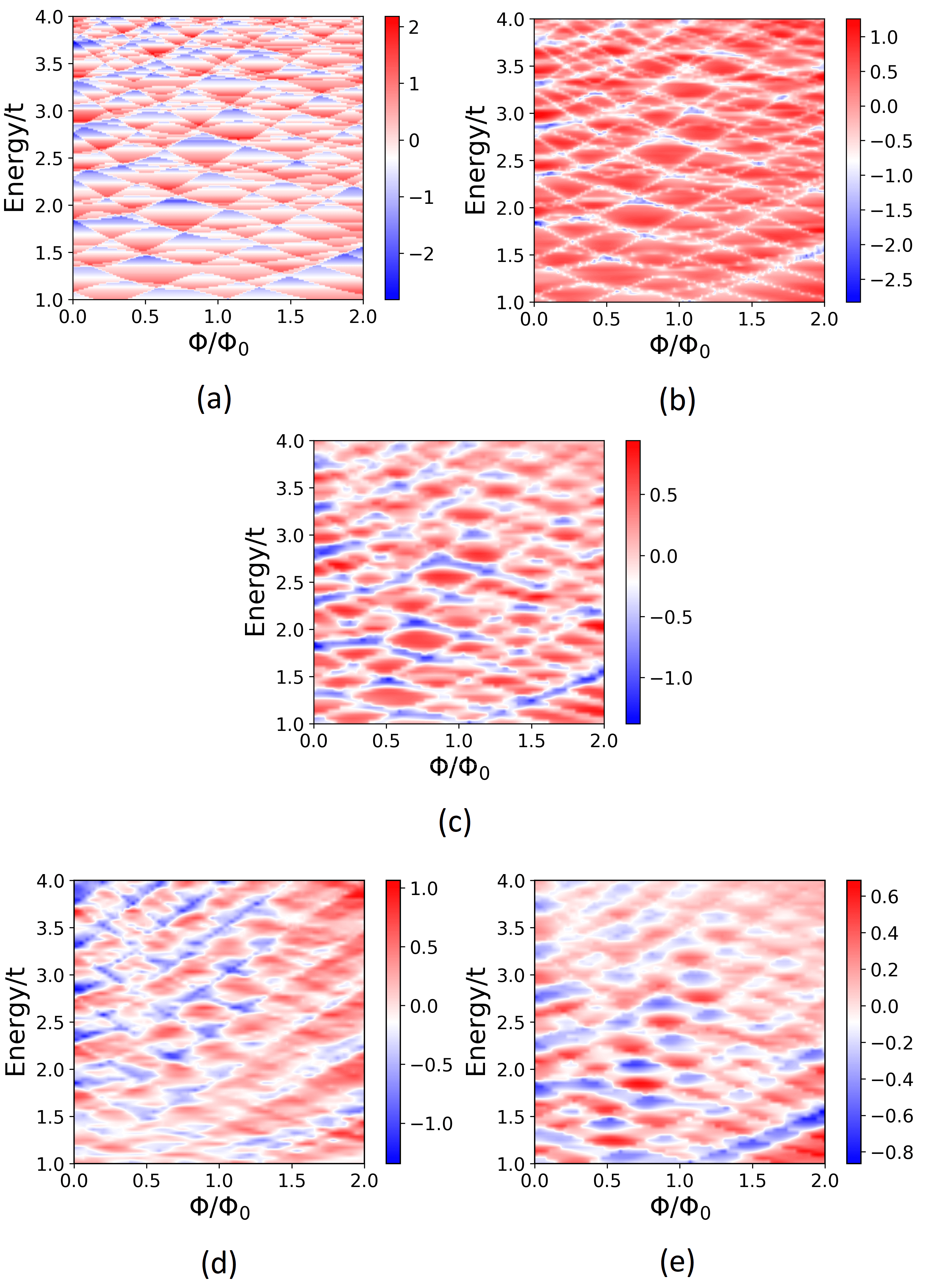}
\caption{$\delta T(E,\Phi)=T(E,\Phi)-\langle T(E,\Phi) \rangle_\Phi$, for each energy for a parabolic transverse potential/weak surface confinement: (a) In the absence of any disorder potential $(W=0)$. (b) In the presence of a spatially uncorrelated disorder with $W=1t$. The angular momentum sub-bands are observed. (c) In the presence spatially uncorrelated disorder with $W=1t$ as well as dephasing characterized by $D=0.4^2$, resulting in smoother variations. (d) In the presence of a spatially Gaussian correlated disorder with standard deviation $\sigma=2a$ and $W=1t$. (e) Same as (d), but with dephasing characterized by $D=0.4^2$. Note that, in all the panels, the conduction band edge (not shown) lies at $\approx -0.1t$.}
\label{T FFT truepar circstep disord}
\end{figure}

Having explored disorder scattering, as well as dephasing in nanowires with a strongly surface confined electronic distribution, it remains to be seen how a weaker surface confinement, like the case shown in Fig.~\ref{T FFT no pot circstep}, affects the results in the presence of disorder. To this end, we study a disordered nanowire of length $25a$ with a parabolic surface confining potential described by~\eqref{tVprop} with $V_0=0.4098t$ and $p=2$, as shown in Fig.~\ref{T FFT truepar circstep disord}, to qualitatively investigate the underlying physics.

On comparing Fig.~\ref{T FFT truepar circstep disord} with Fig. 3 in Ref.~\citen{PhysRevB.91.045422}, which shows the magneto-conductance spectrum in a InAs nanowire, one makes three observations:
 
First, the angular momentum sub-band structure is clearly observed from both our results and the experimental data in Ref.~\citen{PhysRevB.91.045422}. Note that we show a larger range of energy and consequently, many sub-bands are visible. 

Second, as compared to Fig.~\ref{T FFT truepar circstep disord}(b), the experimental data in Ref.~\citen{PhysRevB.91.045422} displays much smoother fluctuations in the magneto-conductance spectrum along with fluctuating behavior within each transmission diamond. Further, the fluctuations are no longer limited to the transmission step edges, but it extends into the transmission plateaus too. This may arise from two sources namely, correlated disorder and dephasing. Now, as seen from~\eqref{lscatt1}, the scattering rate for an uncorrelated disorder potential is proportional to the density of states (detailed in Appendix~\ref{scattlsec}). Accordingly, the scattering rate is peaked and much larger at the Van-Hove singularities than elsewhere. Hence, the effect of disorder is largely limited to band-edges. However, if a disorder potential with a long range real-space correlation ($\langle V(\mathbf{r})V(\mathbf{r}')\rangle \neq \delta(\mathbf{r}-\mathbf{r}')$) is introduced, then there would be contributions from propagators at different wavevectors weighted by the corresponding momentum-space correlation function ($\langle V(\mathbf{q})V(\mathbf{q}')\rangle \neq \delta(\mathbf{q}+\mathbf{q}')$) in the self-energy given by~\eqref{sigborn}, which would then enter the scattering time via~\eqref{tauscatt}. As a result, even for wavevectors not located at the Van-Hove singularities, the scattering rate picks up contributions from the nearby Van-Hove singularities. Therefore, a disorder potential with long-range real-space correlation introduces larger fluctuations within the transmission plateaus compared to uncorrelated disorder. This is evident from comparing the spatially uncorrelated disorder in Fig.~\ref{T FFT truepar circstep disord}(b) with Fig.~\ref{T FFT truepar circstep disord}(d) where the disorder potential at each point is still $\in[-\frac{W}{2},\frac{W}{2}]$ where $W=1t$, but along with a Gaussian spatial correlation with standard deviation $\sigma=2a$, i.e., $\langle V(\mathbf{r})V(\mathbf{r}')\rangle =a^2\frac{W^2}{12}\mathrm{exp}\left(\frac{|\mathbf{r}-\mathbf{r}'|^2}{2\times (2a)^2}\right)$. For disorder potentials with the same strength, the Gaussian correlated disorder yields more fluctuations, in particular, within the transmission plateaus. Note that while this analysis is valid for weak disorder, the conclusion holds even for strong disorder where, at each order of perturbation in $V$, the disorder with long-range real-space correlation yields more fluctuations. Additionally, dephasing can introduce smoother fluctuations, especially within the transmission plateaus. This suggests an interplay of correlated disorder and dephasing in the experiment, both of which must be included in an accurate description of the experiment~\cite{PhysRevB.91.045422}. 

Third, oscillatory features, including even the AB oscillations, seem smeared out in the experiment. This may be attributed to a significant disorder potential. In fact, Fig.~\ref{T FFT truepar circstep disord}(e), with a Gaussian spatially-correlated disorder with standard-deviation $2a$ shows the best qualitative agreement with the experiment. AAS oscillations are not observed in the simulations as our nanowires are shorter than the ones considered in Figs.~\ref{T AAS ALL} and~\ref{T AAS ALL 1} even though a weaker surface confinement renders the observation of AAS oscillations more difficult. In spite of this, our short nanowires qualitatively reproduce the experimentally observed features reasonably well which too do not have the AAS oscillations. This may partly be attributed to the nanowire not being sufficiently long for the transverse potential present, or the presence of significant dephasing.

\section{Conclusion}
We have employed the NEGF formalism to systematically analyze magneto-conductance oscillations in nanowires in the presence of an axial magnetic field, demonstrating the effects of disorder, roughness and dephasing. In the ballistic limit AB oscillations are dominant contingent upon the surface confinement of the electronic distribution in the nanowire. By studying disordered nanowires of different lengths, we show the parameter space which leads to a significant AAS harmonic. We also demonstrated that the relative magnitudes of the scattering length and the device dimensions dictate the ballistic and quantum diffusive regimes within the energy bands, thereby determining the relative dominance of the AB and the AAS oscillations with energy. We find that the AAS oscillations begin dominating from the center of the band, while typical experiments probe the low-energy physics near the conduction band-edge (bottom of the band). This should effectively increase the required disorder strength and/or nanowire length required to see a significant AAS contribution. We then showed the ways of suppressing the oscillatory WL corrections to reveal the non-oscillatory WL correction. Lastly the effects of surface roughness and dephasing on the magnitude as well as the components of the oscillations were studied, revealing a key difference in their effects on the harmonics. While surface roughness may have contrasting effects of dominant AAS oscillations or magnetic depopulation depending on the sensitivity of the hopping parameters to the roughness, dephasing systematically degrades harmonics, beginning with the higher ones. These additional factors can explain the unexpected suppression of the AAS content and the consequent relative dominance of AB oscillations even in disordered nanowires~\cite{PhysRevB.91.045422}. Finally, we considered nanowires with a parabolic transverse potential, demonstrating the necessity of spatially-correlated disorder potential and dephasing to yield qualitative agreement with magneto-conductance experiments in [Holloway et al, PRB 91, 045422 (2015)]. In conclusion, our comprehensive results capture the physics and satisfactorily provide qualitative agreement with experimental features while motivating further experimental research.

{\it{Acknowledgments:}} The authors AL and BM acknowledge support from IIT Bombay SEED grant and ISRO-RESPOND grant. KG and JB acknowledge support from the Natural Sciences and Engineering Research Council of Canada. AL acknowledges useful discussions with Arnab Manna.

Materials supporting the claims shall be made available on reasonable requests.
\appendix
\section{\label{sec:level2} Contact self energies}
For clean nanowires, to reduce computational complexity, only 1 layer (cross section) is taken as the device. Rest of it is accounted for by the contacts~\cite{Golizadeh-Mojarad2008}. To simulate a long nanowire subject to an axial magnetic field, while having a single layer as the device, leads need to have the same geometry and the same magnetic field as the device.

We find the surface Green's function iteratively to calculate the contact self-energies. One may keep the leads free from magnetic field. But, in order to keep the magnetic field divergence-free, it would necessitate either the presence of additional transverse field components to compensate for the axial field gradient, or the use of a gradual ramping (and consequently a much longer device, increasing computational expense) to approximately ensure zero divergence. The surface Green's function $(g_s)$ is given by,
\begin{align}
\beta^{\dagger}g_s\beta-\left(E+i\eta-H_{\mathrm{lead}}\right)+g_s^{-1}=0, \quad \Sigma_{\mathrm{lead}}=\tau g_s \tau^{\dagger},
\end{align}
where $\beta$ is the coupling between the transverse layers in the device and $\tau$ is the coupling matrix between the device and the lead. In our case, $\tau =\beta$.

For a disordered nanowire, the self-energy is calculated in the same way as the ballistic case. However, in this case, the device has a finite length/number of planes to model a disordered nanowire of the corresponding length connected to ideal leads. Note that we have kept the same axial magnetic field in the leads as it cannot be abruptly terminated $(\nabla\cdot \vec{B}=0)$. 
\label{selfE}
\section{Density of states and group velocity of rolled 2D square lattice (cylindrical)}
Assuming spin degeneracy, a band's contribution to the density of states (DOS) is given by~\cite{PhysRevLett.81.2506}, 
\begin{equation}
\mathrm{DOS}(\mathrm{E}) = \frac{2}{l}\sum_i\int dk \delta(k-k_i) \Bigr|\frac{\partial \epsilon}{\partial k_{\parallel}} \Bigr|^{-1},
\label{dosband}
\end{equation}
where $l=2\pi/a$ is the length of the first Brillouin zone, $\epsilon(\mathbf{k})$ is the dispersion relation, and $k_is$ satisfy $E=\epsilon(k_i)$. We follow a procedure, similar to the one given in Ref.~\cite{PhysRevLett.81.2506}. However, for our problem, we cannot use a low energy approximation. We need the distribution over the entire band.
\begin{equation}
\epsilon(\mathbf{k}) = \alpha+2\beta \mathrm{cos}(k_{\parallel}a)+2\beta \mathrm{cos}(k_{\perp}a) \quad (\mathbf{k}=\mathbf{k_{\parallel}}+\mathbf{k_{\perp}}).
\end{equation}
Defining a circumferential vector $\vec{R} = N\vec{a_1}$, where $\vec{a_1}$, is the reciprocal lattice basis vector along the circumference, we get,
\begin{equation}
\Delta k_{\perp}=\Bigr|\mathbf{k}\cdot\frac{\vec{R}}{R}\Bigr|.
\end{equation}
The total DOS is obtained by summing up the sub-band contributions $\mathrm{DOS}(E,n)$.
\begin{align}
\mathrm{DOS}(E) &= \sum_{n=0}^{N-1}\mathrm{DOS}(E,n)\\ 
&=\sum_{n=0}^{N-1}\frac{1}{\pi}\frac{1}{\sqrt{(2\beta)^2-\left( E-\alpha-2\beta \mathrm{cos}(\frac{2\pi n}{N})\right)^2}}. \label{dosem}
\end{align}
Also the group velocity, for each band is given by
\begin{align}
\hbar\mathopen| v(E,n)\mathclose| &= \Biggr|\frac{1}{\hbar}\vec{\nabla}_{\mathbf{k_{\parallel}}}\epsilon(\mathbf{k},n)\Biggr|=(2\beta a)\mathrm{sin}^2(k_{\parallel}a)\nonumber \\
&=a\Bigg( (2\beta)^2-\bigg[ E-\alpha-2\beta \mathrm{cos}\left(\frac{2\pi n}{N} \right) \bigg]^2 \Bigg) ^{(1/2)}. \label{vevm} 
\end{align}
\section{Scattering length}
\label{scattlsec}
Here we derive the scattering length using disorder averaging~\cite{PhysRevB.87.144202}. A random on-site potential (uniformly drawn from $\left[-\frac{W}{2},\frac{W}{2}\right]$) with the following properties is considered. 
\begin{equation}
\langle V(\mathbf{r})\rangle=0 \qquad \langle V(\mathbf{r})V(\mathbf{r}')\rangle=a^2\frac{W^2}{12}\delta(\mathbf{r}-\mathbf{r}').
\label{Vprop}
\end{equation}
where the $\langle\ldots\rangle$ stands for disorder average. From the theory of disorder averaging, the first order term in the self energy, $\Sigma^{(1)}=\int \frac{d\mathbf{q_1}}{(2\pi)^2}\langle V(\mathbf{q_1})\rangle $ vanishes (using~\eqref{tVprop}). The second order term is,
\begin{align}
\Sigma^{(2)}(E, \mathbf{k}) &= \int \frac{d\mathbf{q_1}}{(2\pi)^2}\frac{d\mathbf{q_2}}{(2\pi)^2}\langle V(\mathbf{q_1})G_0(\mathbf{k}+\mathbf{q_2},E)V(\mathbf{q_2}) \rangle  \nonumber\\
&=a^2W^2\int d\epsilon \hspace{0.1cm} \mathrm{DOS}(\epsilon)G_0(E,\epsilon)
\label{sigborn}
\end{align}
For weak disorder, we can use the first Born approximation, in which all but the second order term of the full diagrammatic perturbative expansion are discarded. The imaginary part of the self energy, obtained by using the Sokhotski-Plemelj formula, gives the scattering rate,
\begin{align}
\frac{1}{\tau_{sc}(E)}&=-\frac{2}{\hbar}\mathbb{I}\mathrm{m}\Sigma(E)=\frac{2\pi}{\hbar}a^2\frac{W^2}{12}\mathrm{DOS}(E)\label{tauscatt},\\
l_e(E) &= v(E)\tau_{sc}(E) = \frac{\hbar v(E)}{2\pi}\frac{1}{a^2\frac{W^2}{12}\mathrm{DOS}(E)}.
\end{align}
Using Eqs.~\eqref{dosem} and~\eqref{vevm}, the scattering length at energy E for the $n^{th}$ sub-band is,
\begin{align}
\frac{l_e(E,n)}{a}&= \frac{12}{2\pi W^2}\frac{\mathrm{M}(E,n)}{\mathrm{DOS}(E,n)},\label{lscatt}\\
\mathrm{M}(E,n)&= \Bigg( (2\beta)^2-\bigg[ E-\alpha-2\beta \mathrm{cos}\left(\frac{2\pi n}{N} \right) \bigg]^2\Bigg)^\frac{1}{2},\\
\mathrm{DOS}(E,n)&= \frac{1}{\sqrt{(2\beta)^2-\left( E-\alpha-2\beta \mathrm{cos}(\frac{2\pi n}{N})\right)^2}}. 
\end{align}
The scattering rates are then summed up over all sub-bands~\cite{nemec} using~\eqref{lscatt},
\begin{align}
l_e(E)^{-1} = \frac{1}{N}\sum_{n=1}^Nl_e(E,n)^{-1}.
\end{align}
\label{lscat_deriv}
\section{Parameters}
All energies are specified in units of the hopping parameter $t=\hbar^2/(2ma^2)=0.61$eV, where $m=9.1\times10^{-31}$kg is the electronic mass, and $a=0.79$nm is the lattice constant. This scaling renders the actual value of $t$ seemingly irrelevant, instead manifesting its importance through the length of the nanowire and the realization of actual disorder strength relative to $t$. These are available from experimental data. Our clean nanowires have a diameter equal to $11a$, and the disordered nanowires have diameter equal to $10a$. Disorder potentials, surface roughness parameters and dephasing strengths have been specified in the corresponding figure captions.
\bibliographystyle{apsrev}
\bibliography{bibl}
\end{document}